\documentclass[11pt,letterpaper]{article}
\usepackage{lineno}
\usepackage[margin=1in]{geometry}
\usepackage[T1]{fontenc}
\usepackage{lmodern,todonotes}
\usepackage{amsmath,amssymb,amsthm,enumerate,enumitem}
\usepackage{thmtools,thm-restate}
\makeatletter
\@ifundefined{newcounteralias}{}{%
  \renewcommand\thmt@autorefsetup{%
    \@xa\def\csname\thmt@envname autorefname\@xa\endcsname
      \@xa{\thmt@thmname}%
  }%
}
\makeatother
\usepackage{graphicx}
\usepackage[hidelinks]{hyperref}
\usepackage[nameinlink,capitalise,noabbrev]{cleveref}
\usepackage[ruled,linesnumbered,vlined]{algorithm2e}
\usepackage{multicol, multirow}
\usepackage{arydshln}
\definecolor{forestgreen}{rgb}{0.13, 0.55, 0.13}

\SetCommentSty{mycommfont}
 \setlist[enumerate,itemize]{topsep=4pt,itemsep=2pt,parsep=0pt,partopsep=0pt}
\newtheorem{theorem}{Theorem}[section]
\newtheorem{lemma}[theorem]{Lemma}
\newtheorem{property}[theorem]{Property}

\theoremstyle{definition}
\newtheorem{definition}[theorem]{Definition}

\theoremstyle{remark}

\newcommand{\dist}{\mathop{\mathrm{dist}}\nolimits}

\newcommand{\opt}{\ensuremath{\textsf{OPT}}}
\newcommand{\eps}{\ensuremath{\varepsilon}}
\newcommand{\rt}{\ensuremath{\textsc{r}}}

\newcommand{\weight}{\ensuremath{\textsc{weight}}}

\newcommand{\ddim}{\mathrm{ddim}}

\definecolor{BrickRed}{rgb}{.72,0,0}
\def\EMPH#1{\emph{\textcolor{BrickRed}{#1}}}

\title{Constant-Round Massively Parallel Algorithms for\\
Approximate MSTs in Metric Spaces}
\author{Eunjin Oh\thanks{POSTECH, South Korea. Email: eunjin.oh@postech.ac.kr. This work is supported by Institute of Information \& Communications Technology Planning \& Evaluation (IITP) grant funded by the Korea government (MSIT) (No. RS-2024-00440239, Sublinear Scalable Algorithms for Large-Scale Data Analysis) and the National Research Foundation of Korea (NRF) grant funded by the Korea government (MSIT) (No. RS-2024-00358505).} \and Hyeonjun Shin\thanks{POSTECH, South Korea. Email: \texttt{hyeonjun.shin@postech.ac.kr}.}}
\date{}

\begin{document}
\hypersetup{pageanchor=false}

\begin{titlepage}
\maketitle
\thispagestyle{empty}

\begin{abstract}
We develop a unified algorithmic framework for computing  
approximate minimum spanning trees (MSTs) in metric spaces
in the fully scalable Massively Parallel Computation (MPC) model.
Using this framework, we obtain the following results
for any fixed constant $\sigma\in(0,1)$:

\begin{itemize}
    \item  \textbf{General metrics.}
    Given all pairwise distances, we compute a
    $(1+\varepsilon)$-approximate MST in
    $O(\log(1/\varepsilon))$ rounds using
    $n^\sigma$ local space per machine and
    $\widetilde{O}(n^2)$ total space.

    \item \textbf{Euclidean spaces.}
    Given $n$ points in $\mathbb{R}^d$,  
    we compute an $O(1/\sigma^4)$-approximate MST in
    $O(1)$ rounds using
    $(nd)^\sigma$ local space per machine and
    $O(nd+n^{1+\sigma})$ total space.

    \item \textbf{Doubling metrics.}
    Given $n$ points in a metric of doubling dimension
    $\ddim=O(\sigma\log n)$,
    we compute an $O(1/\sigma)$-approximate MST in
    $O(\ddim/\sigma^4)$ rounds using
    $n^\sigma$ local space per machine and
    $\widetilde{O}(n^{1+\sigma})\cdot 2^{O(\ddim)}$
    total space.
\end{itemize}

For any fixed constant $\varepsilon\in(0,1)$,
these results give the first constant-round
MPC algorithms achieving $(1+\eps)$-approximation 
in general metrics  
and $O(1)$-approximation in a high-dimensional Euclidean space, respectively.
They improve upon the
$O(\log\log n+\log(1/\varepsilon))$-round algorithm
for general metrics [SODA'25] and the
$O(\log\log n\cdot\log\log\log n)$-round algorithm
for Euclidean spaces [SODA'24], respectively.
Moreover, our round complexity for general metrics
is asymptotically optimal in its dependence on
$\varepsilon$, matching the
$\Omega(\log(1/\varepsilon))$ lower bound of
[SODA'25] under the 1-vs-2-cycles conjecture.
Our result also answers an open question raised by [STOC'23].
\end{abstract}

\end{titlepage}
\tableofcontents
\thispagestyle{empty}
\clearpage
\setcounter{page}{1}
\hypersetup{pageanchor=true}

\section{Introduction}
\label{sec:intro}
In this paper, we study the \emph{minimum spanning tree} (MST) problem in metric spaces, which is one of the fundamental problems in computer science.  
Given a set of points and their pairwise distances, the goal is to compute a spanning tree of minimum total weight in the complete graph on the point set, where the weight of each edge is the distance between its endpoints.
The MST problem has been extensively studied, giving rise to a rich body of algorithmic techniques. 
A central application is hierarchical clustering: removing the edges of an MST in decreasing order of weight progressively partitions the point set into clusters~\cite{march2010fast, yaroslavtsev2018massively}.
In particular, MST-based clustering has been used to identify structures in astronomical data~\cite{march2010fast, 10.1093/mnras/stt2136} and to cluster high-dimensional datasets in machine learning~\cite{lai2009approximate}.
MSTs also find applications in biology~\cite{trapnell2014pseudo,yang2020glamst, sammut2024predictability} as well as in 
representation learning~\cite{pmlr-v119-hofer20a, pmlr-v119-moor20a, mordacq2026t}.

\medskip 
Such applications often involve processing large data sets. This motivates the use of modern distributed computing frameworks such as MapReduce, Spark, Hadoop, and Dryad~\cite{dean2008mapreduce, dean2004mapreduce, white2012hadoop, isard2007dryad}.
Although their implementations differ, these systems must coordinate computation across many machines, and the resulting communication  overhead can significantly affect algorithmic efficiency.
The \emph{Massively Parallel Computation (MPC)} model, introduced in~\cite{karloff2010model}, 
has become a central theoretical framework for capturing the algorithmic challenges in these systems.
Over the past decade, various algorithmic problems including the MST problem have been extensively studied in the MPC model~\cite{ azarmehr2025massively, assadi2019massively, bhore2026geometrichashingmpcapplications,  coy2022deterministic, ghaffari2019massively, yaroslavtsev2018massively}.

In the MPC model, 
the input of $\EMPH{N}$ words is distributed across $\EMPH{m}$ machines, each with $\EMPH{s}$ words of local memory, 
giving a total memory of $m \cdot s$ words.
Computation proceeds in synchronous rounds of local computation and communication. In each round, each machine can perform unrestricted local computation within its memory bound, and then send and receive $O(s)$ words.
The primary complexity measures of an MPC algorithm are its round complexity, local memory $s$, and the total space $m\cdot s$.
In this work, we consider the \EMPH{fully scalable} regime where $s=N^\sigma$ for an arbitrary fixed constant $\sigma \in (0,1)$.
Our goal is to minimize the number of rounds, ideally achieving $O(1)$ rounds, while keeping the total space as small as possible.

\medskip 
We focus on the metric MST problem in the fully scalable MPC model. 
Since a metric induces a complete edge-weighted graph, a natural approach is to apply MPC algorithms for MSTs in general graphs.
For general edge-weighted graphs, an \emph{exact} MST can be computed in $O(\log n)$ rounds in the fully scalable regime~\cite{andoni2018parallel, behnezhad2019near, coy2022deterministic}.
However, under the 1-vs-2-cycles conjecture, even computing an \emph{approximate} MST in a general edge-weighted graph requires $\Omega(\log n)$ rounds.\footnote{Distinguishing a single cycle on $n$ vertices from two disjoint cycles on $n/2$ vertices each is conjectured to require $\Omega(\log n)$ rounds in the fully scalable regime. Several conditional hardness results rely on this conjecture~\cite{nanongkai2022equivalence}.}
This barrier has motivated extensive work on MPC algorithms for MSTs in more structured settings, particularly metric spaces, including Euclidean and doubling metrics~\cite{andoni2014parallel,gan2025EMST,chen2021EMD,azarmehr2025massively,jayaram2024massively}. A natural question in this direction is whether approximate MSTs can be computed in $o(\log n)$ rounds by exploiting metric structure.

Although Euclidean MST is a special case of metric MST, their input representations differ significantly in the MPC model. Our goal is to keep the total space close to linear in the input size. In the general metric setting, all pairwise distances are explicitly given, resulting in an input size of $N=\Theta(n^2)$. In contrast, in the Euclidean setting, the input consists of the coordinates of $n$ points in $\mathbb{R}^d$, requiring only $\Theta(nd)$ words. In the doubling metric setting, we assume that each point has an $O(1)$-word representation from which pairwise distances can be computed locally, resulting in an input size of $\Theta(n)$. 
Therefore, directly applying MPC algorithms for general metrics to Euclidean and doubling metric spaces may not achieve the desired total space complexity.

\paragraph{Previous work.}
We review the best-known prior MPC results for MSTs in various metric settings. In the following, we sometimes treat $\sigma$ and $\eps$ as fixed constants. Instead, we indicate any hidden dependence on these parameters by subscripts in the $O(\cdot)$ notation. 

For low-dimensional Euclidean spaces, Andoni, Nikolov, Onak, and Yaroslavtsev~\cite{andoni2014parallel} presented a randomized $O_\sigma(1)$-round MPC algorithm that constructs a $(1+\varepsilon)$-approximate MST in expectation using linear total space. This result also extends to doubling metrics
without increasing the round complexity and the total space complexity. Although they state their results for the $O(1)$-dimensional case, their analysis also applies to higher dimensions: $d=O_\varepsilon(\sigma\log n/\log\log n)$ in Euclidean spaces and $\mathrm{ddim}=O_\varepsilon(\sqrt{\sigma\log n})$ in doubling metrics, using $O(n^\sigma)$ local space.\footnote{For the dimension bounds, see Theorems~2.1 and~3.1 for Euclidean spaces and Theorem~6.6 for doubling metrics in~\cite{andoni2014parallel}.}
As long as the dimensions are within these ranges,
the dependence on the dimension in the round complexity and the total space complexity is linear. 

For high-dimensional Euclidean spaces, 
\cite{10.1145/3564246.3585168} presented an MPC algorithm that estimates the MST cost within a factor of $(1+\varepsilon)$ in $O_{\sigma,\eps}(1)$ rounds using $O(nd)$ total space. However, their algorithm does not construct an actual spanning tree. In fact, constructing a constant-factor approximate MST in high-dimensional Euclidean spaces was more challenging; previously, it could be constructed in $O_\sigma(\log n)$ rounds by constructing a sparse Euclidean spanner and then applying an MPC algorithm for MSTs in general graphs. Jayaram, Mirrokni, Narayanan, and Zhong~\cite{jayaram2024massively} broke this $O(\log n)$ barrier by presenting an $O_{\sigma}(\log\log n\cdot\log\log\log n)$-round MPC algorithm that constructs an $O_\sigma(1)$-approximate MST using $O(nd+n^{1+\sigma})$ total space.

For general metric spaces, \cite{azarmehr2025massively} gave a randomized $O_\sigma(\log(1/\varepsilon)+\log\log n)$-round MPC algorithm that constructs a $(1+\varepsilon)$-approximate MST using $\widetilde{O}(n^2)$ total space. The total space is thus near-linear in the input size. 
Under the 1-vs-2-cycles conjecture, they also proved that any MPC algorithm with $n^{1-\Omega(1)}$ local space and polynomial total space requires $\Omega(\log(1/\varepsilon))$ rounds to construct a $(1+\varepsilon)$-approximate MST in a general metric space, for $1/n\le\varepsilon\le1$.

\medskip 
Despite recent progress on MPC algorithms for metric MSTs, the best-known algorithms for constructing approximate MSTs in general metrics and high-dimensional Euclidean spaces still require a super-constant number of rounds for fixed $\varepsilon$. This raises the following question, which was also explicitly posed in~\cite{10.1145/3564246.3585168,jiang2026round}:

\begin{quote}
\emph{Can one construct approximate MSTs in both general metric spaces and high-dimensional Euclidean spaces in $O(1)$ rounds in the fully scalable MPC model, while maintaining the approximation ratios of the previous best-known algorithms?}
\end{quote}

\paragraph{Our contributions.}
We answer this question affirmatively by developing a unified algorithmic framework for constructing approximate MSTs in general metric spaces and high-dimensional Euclidean spaces. 
In particular, we improve the round complexities of the algorithms of~\cite{jayaram2024massively} and of~\cite{azarmehr2025massively} without increasing the total space complexities and the approximation ratios.
This framework also applies to doubling metric spaces. 
Specifically, we prove the following theorems.


\begin{table*}[t]
    \centering
    \renewcommand{\arraystretch}{1.6}
    \resizebox{\textwidth}{!}{
    \begin{tabular}{@{}lcccccc@{}}
        \hline
        Metric & Dimension & Approximation & Rounds & Total space & Tree? & Source \\
        \hline

        \multirow{2}{*}{General}
        & \multirow{2}{*}{---}
        & $1+\varepsilon$
        & $O_\sigma(\log(1/\varepsilon)+\log\log n)$
        & $\widetilde O(n^2)$
        & $\checkmark$
        & \cite{azarmehr2025massively} \\
        \cdashline{3-7}
        &
        & $1+\varepsilon$
        & $O_\sigma (\log(1/\varepsilon))$
        & $\widetilde O(n^2)$
        & $\checkmark$
        & Theorem~\ref{thm:general-metric-mst} \\
        \hline

        \multirow{4}{*}{Euclidean}
        & $d=O(1)$
        & $1+\varepsilon$
        & $O_\sigma(1)$
        & $O_\sigma(n)$
        & $\checkmark$
        & \cite{andoni2014parallel} \\
        \cline{2-7}
        & \multirow{3}{*}{arbitrary $d$}
        & $1+\varepsilon$
        & $O_\sigma((1/\eps)\cdot \log(1/\varepsilon))$
        & $\widetilde{O}(dn)$
        & ---
        & \rule[-3.5ex]{0pt}{8ex}%
          {\renewcommand{\arraystretch}{1}
           \begin{tabular}[c]{@{}c@{}}
               \cite{10.1145/3564246.3585168}\\
           \end{tabular}} \\
        &
        & $O_\sigma(1)$
        & $O_\sigma(\log\log n\cdot\log\log\log n)$
        & $O(nd+n^{1+\sigma})$
        & $\checkmark$
        & \cite{jayaram2024massively} \\
        \cdashline{3-7}
        &
        & $O_\sigma(1)$
        & $O_\sigma(1)$
        & $O(nd+n^{1+\sigma})$
        & $\checkmark$
        & Theorem~\ref{thm:euclidean-mst} \\
        \hline

        \multirow{2}{*}{Doubling}
        & $\mathrm{ddim}=O_\eps(\sqrt{\sigma\log n})$
        & $1+\varepsilon$
        & $O_\sigma(\mathrm{ddim})$
        & $O_\sigma(n\cdot\mathrm{ddim})$
        & $\checkmark$
        & \cite{andoni2014parallel} \\
        \cdashline{3-7}
        & $\mathrm{ddim}=O(\sigma\log n)$
        & $O_\sigma(1)$
        & $O_\sigma(\mathrm{ddim})$
        & $\widetilde O(n^{1+\sigma})\cdot 2^{O(\mathrm{ddim})}$
        & $\checkmark$
        & Theorem~\ref{thm:doubling-mst} \\
        \hline
    \end{tabular}}
    \caption{
        \small
        Comparison of previous and our results.
The local space per machine is $n^\sigma$ for general and
doubling metrics and $(nd)^\sigma$ for Euclidean spaces,
except that~\cite{10.1145/3564246.3585168} uses
$d^{O(1)}n^\sigma$ local space.
The sixth column with label ``Tree?'' indicates whether the algorithms return an actual spanning tree.
    }
    \label{table:summary}
\end{table*}

\begin{restatable}{theorem}{generalMetricMST}
\label{thm:general-metric-mst}
Fix two parameters $\sigma\in(0,1)$ and $\eps\in(0,1)$. There is an $O(\log(1/(\sigma\eps))/\sigma^2)$-round randomized MPC algorithm that, given all pairwise distances of an $n$-point metric $(X,\dist)$, returns a spanning tree of $X$ whose total weight is at most $(1+\eps)$ times the weight of a minimum spanning tree of $(X,\dist)$. The algorithm succeeds with probability at least $1-n^{-\Omega(1)}$ and uses $n^\sigma$ local space per machine and $\widetilde{O}(n^2)$ total space.
\end{restatable}


\begin{restatable}{theorem}{euclideanMST}
\label{thm:euclidean-mst}
Fix a constant $\sigma\in(0,1)$.
There is an $O(1/\sigma^2)$-round randomized MPC algorithm that,
given a set $P$ of $n$ points in $\mathbb{R}^d$,
returns a spanning tree of $P$ whose total weight is
$O(1/\sigma^4)$ times the weight of a minimum spanning tree of $P$.
The algorithm succeeds with probability at least
$1-n^{-\Omega(1)}$ and uses $(dn)^\sigma$ local space
per machine and $O(nd+n^{1+\sigma})$ total space.
\end{restatable}

\begin{restatable}{theorem}{doublingMST}
\label{thm:doubling-mst}
Fix a constant $\sigma\in(0,1)$.
There is an $O(\ddim/\sigma^4)$-round randomized MPC algorithm that,
given a set $P$ of $n$ points in a metric space
with doubling dimension $\ddim$ satisfying
$\sigma\log n\geq\Omega(\ddim)$,
returns a spanning tree of $P$ whose total weight is
$O(1/\sigma)$ times the weight of a minimum spanning tree of $P$.
The algorithm succeeds with probability at least
$1-n^{-\Omega(1)}$ and uses $n^\sigma$ local space
per machine and $\widetilde{O}(n^{1+\sigma})\cdot 2^{O(\ddim)}$
total space.
\end{restatable}

Table~\ref{table:summary} compares our algorithms with previous MPC algorithms for metric MSTs. For general metrics, we remove the additive $O_\sigma(\log\log n)$ term in the round complexity of Azarmehr, Behnezhad, Jayaram, Łącki, Mirrokni, and Zhong~\cite{azarmehr2025massively}. For high-dimensional Euclidean spaces, we improve the $O_\sigma(\log\log n\cdot\log\log\log n)$-round bound of Jayaram, Mirrokni, Narayanan, and Zhong~\cite{jayaram2024massively} to $O_\sigma(1)$ rounds.
For doubling metrics, we extend the dimension range from
$O_\eps(\sqrt{\sigma\log n})$ in~\cite{andoni2014parallel}
to $O(\sigma\log n)$ while retaining
$O_\sigma(\ddim)$ rounds.
On the other hand, the approximation ratio becomes $O_\sigma(1)$ while the approximation ratio of \cite{andoni2014parallel} is $1+\eps$.

\medskip 
As an application, our algorithms also yield MPC algorithms
for the traveling salesman problem (TSP) in metric spaces.
Doubling the weight of an approximate MST immediately gives
an estimate of the optimal TSP cost, with twice the approximation factor.
We show how to compute an approximate TSP tour itself.
The standard method doubles every tree edge,
computes an Euler tour, and shortcuts repeated visits to each point.
However, under the 1-vs-2-cycles conjecture, computing the traversal
order of an arbitrary tree given only its edges requires
$\Omega(\log n)$ rounds in the worst case in the fully scalable MPC
model~\cite{jayaram2024massively}.
We instead use a hierarchical structure of an approximate MST  constructed
during our MST algorithms to compute an Euler tour.
This gives a $(2+\eps)$-approximation in general metric spaces
and an $O_\sigma(1)$-approximation in Euclidean and doubling
metric spaces, with high probability, without asymptotically
increasing the round complexity, local space per machine,
or total space of the corresponding MST algorithms.
For details, see Section~\ref{sec:tsp}.

\paragraph{Estimating the cost vs. Computing an actual tree.}
Beyond estimating the MST cost, our main contribution is to improve the round complexity for constructing approximate MSTs themselves in the MPC model. This distinction matters not only from an algorithmic perspective but also in applications that require the connections between points. For example, in hierarchical clustering, one can remove the edges of an MST (or an approximate MST) in decreasing order of weight to obtain a nested sequence of partitions of the points. The connected components after each removal determine which points belong to the same cluster at that level. An estimate of the MST cost alone does not identify the edges or the resulting clusters. 
Another example is network design. The edges of a spanning tree specify which pairs of sites to connect; an approximate MST provides such connections with a total weight within a provable factor of the minimum.

\paragraph{Conceptual contribution and main insights.}
A by-now-standard framework for estimating MST costs in sublinear models
uses threshold graphs~\cite{chazelle2005approximating}. For a threshold
$t$, two points are adjacent in the threshold graph if their distance is
at most $t$. By Kruskal's algorithm, the number of MST edges longer than
$t$ is exactly one less than the number of connected components of the
threshold graph at $t$. Thus, by counting the connected components at
thresholds $1,(1+\eps),(1+\eps)^2,\ldots$, we can determine how many
MST edges have lengths between consecutive thresholds and use this
information to estimate the MST cost. This approach has led to efficient
algorithms in several sublinear models~\cite{chazelle2005approximating,czumaj2009estimating,czumaj2005approximating,driemel_et_al:LIPIcs.SoCG.2025.42,10.1145/3564246.3585168}.
However, constructing an approximate MST is more challenging, since
knowing the number of MST edges at each scale does not tell us which
edges to include in the tree. Previous MPC algorithms for computing an approximate MST in metrics addressed this issue in this framework, but required a
super-constant number of rounds~\cite{azarmehr2025massively,jayaram2024massively}.

We instead take a different approach based on Prim's algorithm, inspired
by~\cite{patlin2025sublinear,esfandiari2020parallel}.
In particular, \cite{esfandiari2020parallel} presented an $O(1)$-round \emph{adaptive} MPC algorithm for computing an exact MST in a general graph based on Prim's algorithm.\footnote{In the adaptive MPC model, a machine may make adaptive queries to a distributed hash table within a round, whereas communication in the MPC model is received in the next round. Their algorithm uses stopping conditions different from ours and contracts the trees of an auxiliary forest in $O(1)$ AMPC rounds. Thus, it does not face the difficulty of identifying long trees that arises in our MPC setting. }
This suggests a natural question:
Can a Prim-based approach also be used to construct an approximate MST in fewer MPC rounds?
We run truncated
executions of Prim in parallel on small
neighborhoods that fit on individual machines. The resulting forest 
serves as a building block for our algorithm, whose main insights are as
follows.

\begin{itemize}
\item By choosing appropriate stopping conditions, each truncated execution selects only edges belonging to a common MST of the input. Their union therefore forms a forest. However, this forest may contain very large trees or too many small trees.

\item Large trees may span many machines, making it difficult to identify their connected components. We overcome this problem by cutting edges so that each point can identify the root of its resulting tree locally. 

\item Even when the forest contains many small trees, we can use 
sampling to merge them into fewer trees. The stopping conditions of
the Prim executions allow us to bound the expected total weight of
the edges added by this procedure by a constant factor of the MST cost.
\end{itemize}

For general metrics, we contract each tree into a single vertex and
repeat the procedure. For Euclidean and doubling metrics, we instead
retain one representative point per tree and repeat the procedure. In
each phase, the number of vertices decreases by a factor of $n^{\Omega(\sigma)}$, and thus the algorithm terminates in $O(1/\sigma)$ phases. A more detailed technical overview can be found in Section~\ref{sec:overview}.

\section{Preliminaries}
\label{sec:prelim}
Let $(X,\dist)$ be a metric space on $n$ points.
The \EMPH{doubling dimension} of $(X,\dist)$ is defined as
the smallest value $d$ such that every ball of radius $r > 0$
can be covered by at most $2^d$ balls of radius $r/2$.
It is known that the doubling dimension of $\mathbb{R}^d$ is $\Theta(d)$. 
When we speak about doubling metrics, we treat the doubling dimension $\ddim$ as a parameter; we do not assume that it is constant.
For two nonempty subsets $A,B\subseteq X$, we define \EMPH{$\dist(A,B)$}$=\min_{a\in A,b\in B}\dist(a,b)$. 
Note that these distances need not satisfy the triangle inequality.
We sometimes consider the metric $(X,\dist)$ as the complete graph with vertex set $X$ and edge weight $\dist(\cdot,\cdot)$.
For a graph $G=(V,E)$, we use \EMPH{$V(G)$} and \EMPH{$E(G)$} to denote the vertex set and the edge set of $G$, respectively. 
For a finite set $Z$, we use \EMPH{$|Z|$} to denote the cardinality of $Z$. We sometimes abuse notation and 
use \EMPH{$|G|$} to denote the total number of edges and vertices of $G$.
We use \EMPH{$\opt(X)$} to denote the MST cost of $X$. If it is clear from context, we simply write \EMPH{$\opt$}.
Throughout this paper, we assume that the edge weights are pairwise
distinct, and therefore the MST is unique.  If this is not the case, we can break ties using the lexicographical ordering of the endpoints of the edges. 

For $\lambda>0$, the \EMPH{exponential distribution} with rate $\lambda$, denoted by \EMPH{$\operatorname{Exp}(\lambda)$}, is the distribution on $[0,\infty)$ with probability density function $f(t)=\lambda e^{-\lambda t}$ for $t\geq 0$. In this paper, we use the following standard property of exponential random variables. If $X_1,\ldots,X_m$ are independent random variables with $X_i\sim\operatorname{Exp}(\lambda_i)$, then for any $A\subseteq\{1,\ldots,m\}$, the probability that $\operatorname*{arg\,min}_{1\leq i\leq m} X_i$ belongs to $A$ is $\sum_{i\in A}\lambda_i/\sum_{i=1}^m\lambda_i$.

\paragraph{MPC primitives.}  
The following standard primitives can be done in $O(1/\sigma)$ MPC rounds using $n^\sigma$ local space per machine and total space linear in the number of input and output records~\cite{Goodrich1999Sorting,GoodrichSitchinavaZhang2011Sorting}. Here, each record consists of $O(1)$ words, and the number of input and output records is polynomial in $n$.
\begin{itemize}
    \item \textsc{Sorting}: Given records with comparable keys distributed across machines, we can sort the records by their keys and assign each record its rank in the sorted order. If the records are partitioned into sets $S_1,\ldots,S_k$ and each record in $S_j$ is stored together with $j$, we can also compute its rank within $S_j$ according to the same keys. Each rank can be stored together with its record on the machine that originally stores the record.
    \item \textsc{PrefixSum}: Given a sequence $v_1,\ldots,v_t$ distributed across machines, with each value $v_i$ stored together with its index $i$, we can compute the prefix sum $\sum_{j=1}^{i}v_j$ or the prefix minimum $\min_{1\leq j\leq i}v_j$ for every $i$ and store the result together with $v_i$ on its original machine. For a prefix minimum, we can also return an index attaining the minimum together with any $O(1)$-word data associated with that entry. We can also process multiple sequences separately if each value is stored together with the index of its sequence and its position within that sequence.
    \item \textsc{Aggregation}: Given records $(k,v)$ distributed across machines, we can compute the sum, minimum, or maximum of the values $v$ for each key $k$, even if records with the same key are stored on different machines. The result can be stored together with every record with that key. We can also count the number of records with each key.
    \item \textsc{Broadcasting}: Given records $(k,v)$ with one value $v$ for each key $k$, and requests $(k,i)$ distributed across machines, where $i$ is a machine ID, we can send a copy of $v$ to machine $i$ for every request $(k,i)$. We assume that each machine is the destination of at most $O(n^\sigma)$ requests. All copies are included in the output size.
\end{itemize}

\paragraph{Consistent hashing.}
For the Euclidean and the doubling metric cases, we need to compute a small neighborhood of each point in parallel.
For this, 
we use \EMPH{(geometric) consistent hashing}~\cite{jia2005universal}.
A consistent hash $\varphi: X\rightarrow \mathcal{I}$ for an index set $\mathcal I$ induces a partition of $X$ into \EMPH{clusters} of diameter at most $\Delta$ such that for any two points $x,y\in X$, they are contained in the same cluster if and only if $\varphi(x)=\varphi(y)$.

\begin{definition}[{\cite[Definition~1.6]{czumaj2022streaming}, \cite[Definition~1]{bhore2026geometrichashingmpcapplications}}]
    Let $(X,\dist)$ be a metric space, $\mathcal I$ an index set, and $\Delta>0$. For a distribution $\mathcal H$ over mappings $X\to\mathcal I$, let $\text{supp}(\mathcal H)$ denote the set of mappings that can be sampled from $\mathcal H$. A random mapping $\varphi\sim\mathcal H$ is a \EMPH{$(\Gamma,\Lambda)$-hash} with diameter bound $\Delta$ if:
    \begin{enumerate}
        \item \textbf{[Diameter]} every cluster of every $\varphi\in\text{supp}(\mathcal H)$ has diameter at most $\Delta$.
        \item \textbf{[Consistency]} for every $X'\subseteq X$ with diameter at most $\Delta/\Gamma$, we have $\mathbb E_{\varphi\sim\mathcal H}[|\varphi(X')|]\leq\Lambda$.
    \end{enumerate}
\end{definition}

For Euclidean spaces, Czumaj, Filtser, Jiang, Krauthgamer, 
Veselý, and  Yang~\cite{CJFKVY23,czumaj2022streaming} showed that there is a deterministic hash $\varphi:\mathbb R^d\to\mathbb R^d$ whose description requires $\operatorname{poly}(d)$ space. Following \cite{CGJKV24}, we use this hash $\varphi$ so that each machine can evaluate it locally without communication. The geometric aggregation algorithm in \cite{CGJKV24} provides the following lemma. 

\begin{lemma}[\cite{CGJKV24}]\label{lem:euclidean-consistent-hashing} Let $\sigma\in(0,1)$ be a fixed constant, and let $P\subseteq\mathbb R^d$ be a set of $n$ points with $d=O(\log n)$. For any $\Delta>0$, there exists a $(\Gamma,n^{\sigma/2})$-hash $\varphi$ with diameter bound $\Delta$ and $\Gamma=O(1/\sigma)$, whose description and evaluation require $\operatorname{poly}(d)$ space. Given the same hash on every machine, we can compute the sum, minimum, or maximum of values associated with the points in $P\cap\varphi^{-1}(\varphi(B(x,\Delta/(2\Gamma))))$ for every $x\in P$ in $O(1/\sigma)$ rounds using $n^\sigma$ local space per machine and $O(nd+n^{1+\sigma})$ total space. \end{lemma} For doubling metrics, Bhore, Filtser, Oh, Shin, and Singh~\cite{bhore2026geometrichashingmpcapplications} construct a randomized consistent hash from the input points, together with approximate neighborhood hash pairs. For $\Gamma>0$ and a hash $\varphi:P\to\mathcal I$ with diameter bound $\Delta$, a set $\widetilde{\mathcal P}_{\varphi}\subseteq P\times\mathcal I$ is a set of \EMPH{approximate neighborhood hash pairs} if $(x,\varphi(y))\in\widetilde{\mathcal P}_{\varphi}$ whenever $x,y\in P$ and $\dist(x,y)\leq\Delta/\Gamma$, and $\varphi^{-1}(q)\subseteq B(x,2\Delta)$ for every $(x,q)\in\widetilde{\mathcal P}_{\varphi}$. For each $x\in P$, define $A_P(x,\Delta/\Gamma)=\bigcup_{(x,q)\in\widetilde{\mathcal P}_{\varphi}}\varphi^{-1}(q)$. Then $B(x,\Delta/\Gamma)\cap P\subseteq A_P(x,\Delta/\Gamma)\subseteq B(x,2\Delta)\cap P$. Their result gives the following lemma.

\begin{lemma}[\cite{bhore2026geometrichashingmpcapplications}]\label{lem:doubilng-consistent-hashing}
    Let $\sigma\in(0,1)$ be a fixed constant, and let $(P,\dist)$ be an $n$-point metric space with polynomially bounded aspect ratio and doubling dimension $\ddim\leq c\sigma\log n$ for a sufficiently small constant $c>0$. Given $\Delta>0$ and $\Gamma\geq 16$, we can construct a $(\Gamma,2^{O(\ddim/\Gamma)})$-hash $\varphi$ with diameter bound $\Delta$ and a set of approximate neighborhood hash pairs $\widetilde{\mathcal P}_{\varphi}$ of size $O(n\cdot 2^{O(\ddim/\Gamma)})$, with high probability, in $O(\ddim/\sigma^3)$ rounds using $n^\sigma$ local space and $\widetilde O(n)\cdot 2^{O(\ddim/\Gamma)}$ total space.
\end{lemma}

\section{Technical overview}
\label{sec:overview}

In this section, we give an overview of our algorithms for constructing approximate minimum spanning trees for a metric $(X,\dist)$ on $n$ points. 
We first describe the main ideas of our framework, which apply to both general metrics and Euclidean and doubling spaces. 
But we mainly focus on general metrics in this section. 
At the end of this section, we show how to use consistent hashing to implement the framework 
in the Euclidean and doubling settings
using $O(nd+n^{1+\sigma})$ and $\widetilde{O}(n^{1+\sigma})\cdot 2^{O(\ddim)}$ total space, respectively.


Our starting point differs from that of previous work~\cite{jayaram2024massively,azarmehr2025massively}, which uses a threshold-graph-based approach. Motivated by the use of the Prim algorithm for the MST problem in sublinear models~\cite{patlin2025sublinear,esfandiari2020parallel}, we ask whether a Prim-based approach can be used to construct an approximate MST in the MPC model.
For each point $x\in X$, let $S_x$ be the set of the $n^\sigma$ points in $X$ closest to $x$, including $x$. We run the Prim algorithm starting from $x$ on the points in $S_x$ on a single machine. Since this computation ignores points outside $S_x$, it may select edges that do not belong to an MST of $X$. To avoid this, we set $\tau_x := \frac12\max_{y\in S_x}\dist(x,y)$. Every point outside $S_x$ is at distance at least $2\tau_x$ from $x$. We stop the local computation when the total weight of the current tree reaches $\tau_x$ or the next edge has weight at least $\tau_x$. With these choices, the local computation stops before it would select an edge different from the one selected by the Prim algorithm on $X$, and therefore, using the same tie-breaking rule in all runs, all selected edges belong to the same MST of $X$. 
We call this procedure \textsc{TruncatedPrim}.
Let $F$ be the forest on $X$ whose edges are the union of the edges selected from all points $x\in X$.

\paragraph{Does $F$ really help?}
The forest $F$ would be useful if it had at most $n^{1-\sigma}$ components each with $O(n^\sigma)$ vertices, and if each component could be identified and gathered on a single machine in $O(1)$ rounds. In a general metric, we could contract each component into a single vertex; in a Euclidean space, we could instead choose one representative point from each component. Either way, the resulting instance has at most $n^{1-\sigma}$ vertices. If such a reduction decreases the number of vertices by a factor of $n^\sigma$ at every step, we could repeat it for $O(1/\sigma)$ steps to obtain an exact MST in a general metric and an $O(1/\sigma)$-approximate MST in a Euclidean space. 
In particular, there is an instance that realizes this condition. Consider a point set on a line consisting of $n^{1-\sigma}$ groups of $n^\sigma$ equally spaced points each, where consecutive groups are sufficiently far. For each point $x$, the set $S_x$ consists of the points in its group. Then the components of $F$ are precisely the groups. Thus we may safely reduce the instance size in this case.

In general, however, $F$ need not have such a convenient structure. We may encounter two difficulties:
(i) a component is too large to fit on a single machine;
(ii) the number of components is too large to reduce the instance by a polynomial factor. 
More specifically, consider 
the case where all $n$ points are equally spaced on a line.
In this case, $F$ can be a single path on all $n$ points, which is too large to gather on one machine. At the other extreme, if every pair of distinct points in a general metric has distance $1$, then $\tau_x=1/2$ for every $x$. The first candidate edge in each local computation has length $1$, so no edge is selected. In this case, $F$ consists of isolated vertices and does not reduce the number of points.

\paragraph{Resolving Issue (i): Forest cutting\textcolor{red}{.}}
We start with the following observation.
We indeed do not need to gather each component on a single machine for our purposes. Instead, it suffices for each vertex to know an identifier of its component, which can be used to select a representative or to identify vertices to be contracted.
However, identifying the components of a forest is still challenging in the MPC model. Indeed, a constant-round MPC algorithm for identifying the components of an arbitrary forest would contradict the 1-vs-2-cycles conjecture~\cite{DBLP:conf/soda/BalliuLMOU23}. Instead, we cut $F$ into rooted subtrees and identify these subtrees \emph{using local neighborhoods}. More specifically, we first orient and remove some edges to obtain a rooted forest, and then use random labels to cut additional edges. The cuts ensure that, for each vertex $x$, the root of its resulting subtree lies in a small forest neighborhood available on its machine. Thus, each vertex can identify its root locally. 

\paragraph{Resolving Issue (ii): Forest merging.} A simple solution is sampling: we sample each component independently with a probability proportional to its size so that the expected number of sampled components is polynomially smaller than $n$. We connect each unsampled component to its closest sampled component.\footnote{In the Euclidean setting, we instead use an approximate nearest root of a sampled component to determine how to merge the components. This can be implemented efficiently using consistent hashing.} The added edges may not belong to an MST, so we need to bound their weight. For each unsampled component containing a vertex $x$, its neighborhood $S_x$ intersects a sampled component with high probability, so the edge added for this component has weight at most $2\tau_x$. 
To see why the total cost is small, choose a representative with the minimum threshold in each component of $F$. The stopping conditions allow us to charge each such threshold to the weight of its component and an MST edge leaving it.
Since each MST edge is charged at most twice, the sum of these thresholds for all representatives is $O(\opt)$.\footnote{The forest cutting also ensures that the expected sum of the thresholds of the roots after cutting is $O(\opt)$.}
This implies that the expected total weight of the added edges is also $O(\opt)$. 
A more detailed explanation will be provided in Section~\ref{sec:overview-merging}.
This is sufficient for the approximation factor claimed for the Euclidean and doubling cases. 
For general metrics, we first merge components along MST edges before sampling, reducing the expected sum of $\tau_x$ for the representatives $x$ to $\eps\opt$.  

\medskip 
We now explain the ideas behind forest cutting and forest merging in more detail, focusing on the first iteration for a general metric. Later iterations require additional care, as each contracted vertex must carry the weight of the tree it represents. Details can be found in Section~\ref{sec:overview-completing} and Section~\ref{sec:general}.
We also briefly discuss the modifications for Euclidean and doubling metrics in Section~\ref{sec:overview-completing}.

\subsection{Forest cutting (for the first iteration)}
Let $F$ be the forest obtained by running \textsc{TruncatedPrim} from every vertex $x\in X$. The edges of $F$ are distributed across the machines. For each $x\in X$, one machine stores the neighborhood $S_x$ and the subgraph of $F$ induced by $S_x$. Our goal is to remove edges from $F$ so that each vertex can identify the root of its resulting tree using local information, while the expected sum of the thresholds $\tau_x$ over the roots $x$ is $O(\opt)$. 
The forest cutting algorithm consists of two steps. Both steps use the following property, which follows from the definition of $\tau_x$ and the triangle inequality.
\begin{property}\label{prop:overview-thresholds}
For any $x,y\in X$, we have $|\tau_x-\tau_y|\leq \dist(x,y)/2$.
\end{property}

\paragraph{Step 1. Giving orientations to edges.}
We first orient and possibly remove edges to ensure that each vertex has at most one outgoing edge. Each resulting tree then has a unique vertex with no outgoing edge, which we call its root. For each non-root vertex, the endpoint of its outgoing edge is its parent. The parent relation defines a unique path from each vertex to its root. When we remove an edge, its tail becomes the root of a new tree.
We give orientations to edges as follows. For each vertex $x$, choose the vertex $v(x)$ with the lexicographically smallest identifier among the vertices at distance at most $\tau_x/4$ from $x$ in $F$, where distance in $F$ is the total weight of the edges on the path. If $v(x)\ne x$, orient the first edge on the path from $x$ to $v(x)$ towards $v(x)$. 
Since the size of $S_x$ is small, $v(x)$ and the first edge can be computed locally. After all vertices make their choices in $F$, remove every edge that has not been oriented.

To see why each remaining edge has a unique orientation, suppose that an edge $(x,y)$ is oriented in opposite directions. By Property~\ref{prop:overview-thresholds}, $v(x)$ lies within the search region of $y$ and $v(y)$ lies within the search region of $x$. The choice of the smallest identifier then implies $v(x)=v(y)$, but their paths to this vertex cannot start with the same edge in opposite directions,
which is a contradiction.
Each vertex orients at most one edge out of itself and therefore has at most one outgoing edge. A resulting tree with $k$ vertices has $k-1$ edges, whose outgoing endpoints must be distinct. Thus, exactly one vertex has no outgoing edge, and this implies that each tree has a unique root.

For forest merging, we need to show that the sum of $\tau_x$
for the roots is $O(\opt)$. Each root $x$ chooses itself as $v(x)$; therefore, no vertex with a smaller identifier lies within distance $\tau_x/4$ from $x$ in $F$. Together with Property~\ref{prop:overview-thresholds}, this implies that the neighborhoods in $F$ of radius $\tau_x/16$ centered at the roots $x$ are pairwise disjoint. For the analysis, view each edge of $F$ as a line segment of length equal to its weight. For each original component of $F$ before Step 1, consider paths $\pi_x$ from the roots $x$ produced by Step 1 in that component towards a vertex $u$ of minimum threshold in the component, stopping each path from a root $x$ at that vertex or at distance $\tau_x/16$ from $x$, whichever comes first. The stopping point may lie in the interior of an edge. Since these path segments $\pi_x$ are disjoint, their total length is at most the weight of $F$. If $\pi_x$ has length $\tau_x/16$, we charge $\tau_x$ to the length of $\pi_x$. Otherwise, $\pi_x$ reaches $u$, and Property~\ref{prop:overview-thresholds} implies that $\tau_x=O(\tau_u)$. As the paths are disjoint, at most one path in each component reaches $u$; we charge its threshold to $\tau_u$. The weight of $F$ is $O(\opt)$ because $F$ is contained in an MST, and the sum of the minimum thresholds of the original components of $F$ is $O(\opt)$ by the stopping condition of \textsc{TruncatedPrim}.
This implies that the sum of $\tau_x$ for all roots is $O(\opt)$.

\paragraph{Step 2. Cutting additional edges from the oriented forest.}
After Step~1, each non-root vertex can identify its parent in the resulting forest. However, a vertex may not know the root of its tree; we need this information to contract the tree into a single vertex. In Step~2, we cut additional edges to ensure that, for each vertex $x$, the path in $F$ from $x$ to its root lies in $S_x$, while the expected sum of $\tau_x$ over all roots remains $O(\opt)$.
If so, $x$ can identify its root locally using the subgraph stored on its machine. 

To achieve this, we use the following strategy. For technical reasons, we attach an infinite ray to each root of the forest obtained after Step~1. For each vertex $x$, let $\pi_x'$ be the path of length $\tau_x/8$ starting at $x$ and following parent edges towards its root, continuing along the attached ray if necessary. The endpoint of the path may lie in the interior of an edge or on a ray. First, the root of $x$ after cutting must lie on $\pi_x'$. Second, each edge $(v,p(v))$ is cut with probability $O(\dist(v,p(v))/\tau_v)$, where $p(v)$ denotes the parent of $v$ in $F$. The first property allows $x$ to identify its root locally. For the second, cutting $(v,p(v))$ makes $v$ a new root, and its expected contribution to the sum of $\tau_x$ for all roots is $O( \tau_v \cdot \dist(v,p(v))/\tau_v )=O(\dist(v,p(v)))$.

We start with the following observation. For every edge $(v,p(v))$, Property~\ref{prop:overview-thresholds} implies that the total length of the portions belonging to exactly one of $\pi_v'$ and $\pi_{p(v)}'$ is at most $3\dist(v,p(v))$. Since the union of the two paths has length at least $\tau_v/8$, the fraction of the union that belongs to exactly one path is at most $\min\{1,24\dist(v,p(v))/\tau_v\}$. For example, if $\dist(v,p(v))\leq\tau_v/48$, the two paths share at least half of their union.
For each edge $(v,p(v))$, we aim to design a cutting scheme so that the probability of cutting the edge is at most the fraction of $\pi_v'\cup\pi_{p(v)}'$ belonging to exactly one of the two paths, which gives the desired bound of $O(\dist(v,p(v))/\tau_v)$. 

To see how this observation can be useful, focus on the two paths $\pi_v'$ and $\pi_{p(v)}'$. Their union consists of three pieces: the piece common to both paths and one piece belonging only to each path. Suppose that each nonempty piece of length $\ell$ receives an independent random key drawn from $\operatorname{Exp}(\ell)$, and each endpoint chooses the smallest key on its path. Then $v$ and $p(v)$ choose the same key precisely when the smallest key lies in the common piece, which happens with probability equal to the length of the common piece divided by the length of the union. This satisfies the probability bound mentioned earlier.

To coordinate these choices across all vertices, we subdivide the edges of the forest and the rays attached to the roots into \emph{fragments} and assign each fragment $I$ an independent random key drawn from $\operatorname{Exp}(\ell(I))$, where $\ell(I)$ is the length of $I$.  Specifically, the fragments are the finite segments obtained by subdividing each edge and ray at the endpoints of the paths $\pi_x'$ for all vertices $x$. We assign a distinct label to each ray and to each fragment on an edge, and give every fragment on a ray the label of that ray. Each vertex $x$ takes the label of the fragment with the smallest key on $\pi_x'$. We cut an edge if its endpoints have different labels. By the argument above, the probability that a fixed edge $(v,p(v))$ is cut 
is $O(\dist(v,p(v))/\tau_v)$, as desired. Moreover, for each vertex $x$, let $I$ be the fragment with the minimum key in the path $\pi_x'$. If $I$ lies on a ray, the root of $x$ before Step~2 lies on $\pi_x'$. Otherwise, let $(u,p(u))$ be the edge containing $I$. Since $I$ has a unique label, $p(u)$ cannot have the same label as $x$. Thus, we cut an edge on the path from $x$ to $p(u)$. In either case, the root of $x$ lies on $\pi_x'$, and every vertex on the path from $x$ to this root belongs to $S_x$.

\subsection{Forest merging (for the first iteration)}\label{sec:overview-merging}
After forest cutting, we have a rooted forest $F$ such that, for each vertex $x$, every vertex on the path from $x$ to its root belongs to $S_x$.
Moreover, the expected sum of $\tau_z$ for all roots $z$ is $O(\opt)$.
Our goal is to merge the trees of $F$ by adding edges of total length at most $\eps\opt$, leaving at most $n^{1-\sigma}$ components. 
As discussed earlier in this section, we can use sampling to merge the trees. However, the expected cost of the edges added by this approach is bounded in terms of the expected sum of $\tau_z$ over all roots $z$, which gives only an $O(\opt)$ additive error. This bound is too large for our purpose, as we aim for an additive cost of at most $\eps\opt$. Before sampling, we therefore merge the trees using a different strategy to reduce the sum of $\tau_z$ over their roots.

\paragraph{Merging via leader compression.}   We first merge the trees of $F$ along MST edges without introducing any additive error. In each iteration, every component selects a minimum-weight edge leaving it.  The cut property ensures that these edges are contained in an MST. 
Instead of adding all these edges, we use leader compression to merge components while keeping track of the resulting components in the MPC model.  Each component independently becomes a leader with probability $1/2$. For each nonleader, if its selected edge connects it to a leader, we merge the two components along that edge. The root of each merged component is the root of its leader. 
A similar combination of a Borůvka-style algorithm and leader compression was also used in \cite{azarmehr2025massively} to reduce the number of components. Here, we use this approach to reduce the expected sum of $\tau_z$ over the remaining roots. In each iteration, each root is removed with constant probability, and the expected sum of the thresholds of the remaining roots decreases by a constant fraction. After $O(\log(1/\eps))$ iterations, this sum is at most $\eps\opt$ in expectation.

\paragraph{Merging via sampling.}  Now we merge the remaining components by sampling. This is the only step that can introduce an additive error. We sample each component $C$ independently with probability $\min\{1,|C|/n^{\sigma/10}\}$.  For each unsampled component, we add a shortest edge connecting it to a sampled component. The expected number of sampled components is at most $n^{1-\sigma/10}$, and every unsampled component is connected to one of them. For each root $z$, the components intersecting $S_z$ contain at least $n^\sigma$ points in total. By the sampling probabilities, at least one of these components is sampled with high probability.  Since every point in $S_z$ is within distance $2\tau_z$ of $z$, the edge added for an unsampled component with root $z$ has length at most $2\tau_z$. Thus, on this high-probability event, the total length of the added edges is at most twice the sum of $\tau_z$ over the roots before sampling. By the preceding step, this gives an $O(\eps\opt)$ bound on the expected contribution from this event; the contribution from the remaining low-probability outcomes can be bounded separately. Consequently, this step reduces the expected number of components to $O(n^{1-\sigma/10})$ while adding edges of total expected length $O(\eps\opt)$.

\subsection{Completing the algorithms}\label{sec:overview-completing}
We now explain how to extend the first iteration to obtain an approximate MST in a general metric. We contract each component of the forest into a single vertex.  We would like to repeat the same procedure on the contracted graph, using the minimum distance between each pair of components as their edge weight. However, 
the resulting graph need not satisfy the triangle inequality.  
To resolve this, we introduce vertex weights: 
For each contracted vertex $v$, let $C_v$ denote its component and let $\omega_v$ be the total weight of its spanning tree. We define the distance between two vertices $u$ and $v$ by $\dist(u,v):=\min_{a\in C_u,\,b\in C_v}\dist(a,b)$, where the distances on the right are measured in the original metric. These distances and vertex weights satisfy the relaxed triangle inequality: $\dist(u,v)\le \dist(u,w)+\omega_w+\dist(w,v)$. Indeed, the two points of $C_w$ realizing the distances to $C_u$ and $C_v$ can be connected by a path in its spanning tree of length at most $\omega_w$. This inequality allows us to extend the algorithm to the instances produced in later iterations. 

In this way, each iteration reduces the number of vertices by a polynomial factor with high probability. After $O(1/\sigma)$ iterations, the remaining graph fits on one machine, where we compute an exact MST. By setting the error parameter of each iteration to $O(\sigma\eps)$, we obtain a $(1+\eps)$-approximate MST. Most operations in each iteration, including computing the neighborhoods directly, can be implemented in $O(1/\sigma)$ MPC rounds using $O(n^2)$ total space. The repeated leader compression takes $O(\log(1/(\sigma\eps))/\sigma)$ rounds per iteration. Overall, the algorithm takes $O(\log(1/(\sigma\eps))/\sigma^2)$ rounds and uses $n^\sigma$ local space per machine and $\widetilde{O}(n^2)$ total space.

\paragraph{Modification for Euclidean and doubling spaces.}
For Euclidean and doubling spaces, there are two key differences. First, contracting components need not preserve the underlying metric structure. We instead select one representative from each component and continue with these points. Since the representatives form a subset of the original points, their MST has weight at most twice that of an MST on the original points. This allows us to bound the weight added at each iteration in terms of the original MST weight and obtain an $O_\sigma(1)$-factor approximation.

Second, most operations can be implemented in $O_{\sigma}(1)$ MPC rounds using $O(n^{1+\sigma})$ total space, apart from the space needed to store the input points. Computing the \emph{exact} neighborhoods $S_x$ directly, however, involves all pairwise distances and uses quadratic space. We use consistent hashing to compute \emph{approximate} neighborhoods and \emph{approximate} nearest sampled roots without constructing all pairwise distances. However, the thresholds derived from approximate neighborhoods may not satisfy Property~\ref{prop:overview-thresholds}, which is crucial for forest cutting. A key technical step is to smooth the approximate thresholds locally, yielding thresholds that satisfy this property while remaining comparable to the exact thresholds.
In this way, we show how to retain a constant approximation factor for fixed $\sigma$. The resulting total space is $O(nd+n^{1+\sigma})$ for points in $\mathbb{R}^d$ and $\widetilde{O}(n^{1+\sigma})2^{O(\mathrm{ddim})}$ for metrics of doubling dimension $\mathrm{ddim}$.

\paragraph{Computing a TSP tour.} As mentioned earlier, we can obtain a TSP tour by doubling the edges of our approximate MST, computing an Euler tour, and shortcutting repeated vertices. The challenge is to compute the Euler tour efficiently in the MPC model. After forest cutting, each root-to-vertex path fits on a single machine, which allows us to compute an Euler tour of each tree in parallel.  We then combine these tours as the trees are merged, following the hierarchy of the MST construction and using the algorithm of~\cite{jayaram2024massively}. This yields a TSP tour of weight at most twice the weight of our tree without asymptotically increasing the round or space complexity.

\paragraph{Roadmap.}
The algorithms for general metrics and for Euclidean and doubling metrics share the same framework, with the modifications described above. In Section~\ref{sec:subroutine}, we present truncated Prim and forest cutting as common subroutines with parameters that accommodate both settings. We present the general metric algorithm in Section~\ref{sec:general} and the algorithms using consistent hashing in Section~\ref{sec:euclidean}. Finally, we describe the construction of TSP tours in Section~\ref{sec:tsp}.

\section{Key ingredients: Truncated Prim and forest cutting}
\label{sec:subroutine}
In this section, we introduce two subroutines, \textsc{TruncatedPrim} and \textsc{ForestCutting}.
In our main algorithms, the input to each call to \textsc{TruncatedPrim} fits in a single machine, so we always run \textsc{TruncatedPrim} locally. 
The input to \textsc{ForestCutting} is a
forest with $n$ vertices; we do not assume any underlying metric.
Here, we provide an MPC implementation using $O(N)$ total space, where $N$ is the input size defined below. These subroutines will be used in both the Euclidean and general metric cases.

\subsection{Truncated Prim's algorithm}
In this subsection, we introduce a subroutine, which we call \textsc{TruncatedPrim},
that constructs a subgraph of an MST without computing the entire MST.\footnote{Although \cite{esfandiari2020parallel} also used the truncated Prim algorithm, their stopping conditions are different from ours.}
Throughout this subsection, we use $X$ for the vertex set of the complete graph to which the subroutine is applied, and $\dist$ for its edge lengths. Recall that Prim's algorithm starts from a vertex $x\in X$
and grows a tree one edge at a time.
Let $U$ denote the set of vertices reached so far,
initially $U=\{x\}$.
At each iteration, we select an edge $uv$
minimizing $\dist(u,v)$ among all edges with
$u\in U$ and $v\in X\setminus U$.
We then add this edge to the current tree and add $v$ to $U$.
In our case, we run Prim's algorithm on a small neighborhood $S_x$ of $x$ that fits in a single machine. Here, an issue is that if a vertex $y$ with $y\notin S_x$ has a short edge to a vertex in $S_x$, an MST of $S_x$ obtained by Prim's algorithm starting from $x$ is not necessarily 
contained in an MST of $X$. To resolve this issue, we choose a neighborhood and a (sufficiently small) \emph{threshold} together to terminate the procedure before we visit a vertex having a short edge to a vertex in $X\setminus S_x$. 

More generally, we use the following setting. 
Each vertex $x\in X$ is associated with a positive threshold \EMPH{$\tau_x$}, a neighborhood \EMPH{$S_x$} with $x\in S_x\subseteq X$, and a nonnegative vertex weight \EMPH{$\omega_x$}. We denote a call to this subroutine by \textsc{TruncatedPrim}$(x,S_x,\tau_x,\omega)$, where $\omega$ denotes the vertex weights $\{\omega_y\}_{y\in S_x}$. Then the algorithm works as follows. For pseudocode, see 
Algorithm~\ref{alg:truncated-prim}.
Let $U$ be the set of vertices visited so far, and $E_x$ be the set of chosen edges.
Thus the set of the endpoints of the edges of $E_x$ coincides with $U$ unless $E_x=\emptyset$. 
We use \EMPH{$\omega(E_x)$} to denote the total length of $E_x$ plus $\sum_{v\in U}\omega_v$. 
We terminate the procedure when
$\omega(E_x) \geq \tau_x$ or $U=S_x$.  
At each iteration, we choose an edge $(u,v)$
with $u\in U$ and $v\in S_x\setminus U$
minimizing $\dist(u,v)$. 
Then we add $(u,v)$ to $E_x$ if its length is less than $\tau_x$; otherwise, we immediately terminate the procedure. Then we add $v$ to $U$, and $\omega(E_x)$ increases by $\dist(u,v)+\omega_v$ accordingly.

\begin{algorithm}[htbp]
\caption{\textsc{TruncatedPrim}$(x,S_x,\tau_x,\omega)$}\label{alg:truncated-prim}
\small
\SetKwInput{KwIn}{Input}
\SetKwInput{KwOut}{Output}
\KwIn{Root $x$, neighborhood $S_x\ni x$, threshold $\tau_x>0$,
  and vertex weights $\{\omega_y\}_{y\in S_x}$}
$U\gets\{x\}$; $E_x\gets\varnothing$\;
\While{$\omega(E_x)<\tau_x$ \textnormal{and} $U\ne S_x$}{
  Let $(u,v)$ be an edge with $u\in U$ and $v\in S_x\setminus U$ minimizing $\dist(u,v)$\;
  \If{$\dist(u,v)\ge\tau_x$}{
    \textbf{break}\;
  }
  $E_x\gets E_x\cup\{\{u,v\}\}$; $U\gets U\cup\{v\}$\;
}
\Return{$E_x$}\;
\end{algorithm}

In Section~\ref{sec:euclidean} for the Euclidean case, we always apply
Algorithm~\ref{alg:truncated-prim} by setting $X$ as a point set in $\mathbb{R}^d$ and $\dist$ as the Euclidean distance in $\mathbb{R}^d$. We set $\omega_x=0$ for all $x\in X$.
For general metrics, however,
we contract several vertices into a single vertex before
applying Algorithm~\ref{alg:truncated-prim}; see Section~\ref{sec:general}. 
Here, each vertex $x$ represents a set $C_x$ of vertices of the original metric,
and we abuse the notation to denote  $\dist(x,y):=\dist(C_x,C_y)$.
Thus $\dist(\cdot,\cdot)$ does not satisfy the triangle inequality in general. 
Instead, by setting the vertex weight $\omega_z$ to the total weight of the tree maintained within $C_z$, we will prove the following inequality:
\begin{equation}
\label{eqn:triangle}
\dist(u,v)\le \dist(u,z)+\omega_z+\dist(z,v)
\qquad\text{for all }u,v,z\in X.
\end{equation}
This inequality also holds for the Euclidean case where we set $\omega_z:=0$ and $\dist(u,z):=\|u-z\|$. Therefore, in this subsection, we assume that $\dist(\cdot,\cdot)$ satisfies Equation~(\ref{eqn:triangle}).

\paragraph{Analysis.}
The property we want to achieve is the following:
\textsc{TruncatedPrim} returns a subgraph of an MST of $X$. However, this can be achieved only when the parameters are chosen carefully.
In particular, we show that this can be achieved if
$\omega_x+\dist(x,y)>2\tau_x$ for every $y\notin S_x$. 

Let $T$ be the MST of the complete graph on $X$ with edge lengths $\dist$. We assume that the distance values are pairwise distinct. If this is not the case, we break ties by the lexicographical order of the endpoints, using the same rule throughout. Therefore, $T$ is uniquely determined under this rule.
In particular, if we apply Prim's algorithm from any vertex $x$, the algorithm returns $T$.

\begin{lemma}\label{lem:truncated-prim}
Let $x$ be a vertex in $X$, and let $S_x$ be a neighborhood with $x\in S_x\subseteq X$ such that $\omega_x+\dist(x,y)>2\tau_x$ for every $y\in X\setminus S_x$.
Let $E_x$ be the set of edges returned by \textnormal{\textsc{TruncatedPrim}}$(x,S_x,\tau_x,\omega)$.
Then $E_x\subseteq E(T)$.
\end{lemma}
\begin{proof}
We prove by induction on $k$ that the first $k$ edges chosen by \textsc{TruncatedPrim} coincide, in the same order, with the first $k$ edges chosen by Prim's algorithm on $X$ starting from $x$.
The claim holds immediately for $k=0$.
Suppose that the claim holds after $k$ edges, and that \textsc{TruncatedPrim} is about to add another edge. 
Let $U$ be the set of vertices visited by 
\textsc{TruncatedPrim} (and thus by Prim's algorithm), and $E_x$ the current edge set.

We have $\omega(E_x)<\tau_x$. Let $(u,v)$ be the next edge chosen by Prim's algorithm, with $u\in U$ and $v\in X\setminus U$.
Its length is at most that of the edge added by \textsc{TruncatedPrim}, and hence is less than $\tau_x$.
Then we have the following inequalities:
\[
\omega_x+\dist(x,v)\le \omega(E_x)+\dist(u,v)<2\tau_x.
\]
The first inequality can be obtained by applying Equation~(\ref{eqn:triangle}) along the path from $x$ to $u$ in the current tree.
By the assumption on $S_x$, 
we have $v\in S_x$.
Therefore, \textsc{TruncatedPrim} also chooses $(u,v)$. This implies that $E_x\subseteq E(T)$.
\end{proof}


\subsection{Forest cutting algorithm}
In this subsection, we introduce a subroutine, which we call \textsc{ForestCutting}, that computes a \emph{low-cost forest partition} defined as follows.
Consider an $n$-vertex forest \EMPH{$F$} with positive edge weights \EMPH{$\ell_e$}, non-negative vertex weights \EMPH{$\omega_x$}, and positive vertex costs \EMPH{$a_x$}.
Here, the edge and vertex weights define the distances in $F$, whereas the vertex costs $a_x$ determine the quality of the partition as we will see later.
For distinct vertices $x$ and $y$ in the same component of $F$, let \EMPH{$\dist_F(x,y)$} be the total weight of the edges and vertices on their unique path, including $x$ but excluding $y$.
We set $\dist_F(x,x)=0$.
Notice that $\dist_F(x,y)-\dist_F(y,x)=\omega_x-\omega_y$ if $x$ and $y$ are contained in the same component.
We say that a tree is \EMPH{oriented} if its edges are oriented towards a unique root, and that a forest is oriented if each of its components is oriented.
For a non-root vertex $x$ in an oriented forest, we use \EMPH{$p(x)$} to denote its parent.
Removing the edge $(x,p(x))$ splits its component into two oriented trees, one with the original root and one with the new root $x$.

Our goal is to choose and orient a subset of the edges of $F$ so that the resulting rooted subtrees satisfy the following definition.
The first term in (ii) is a lower bound on the total root cost of any partition: each component of $F$ contains at least one root.

\begin{definition}[Low-cost forest partition]\label{def:low-cost-forest-partition}
A randomized partition of $F$ into oriented subtrees obtained by removing edges and orienting the remaining edges is a \EMPH{low-cost forest partition} if the following properties hold.
Let $R$ be the set of roots of the resulting trees, and let $\rt(x)$ be the root of the tree containing $x$.
\begin{enumerate}
\renewcommand{\labelenumi}{(\roman{enumi})}
\item For every vertex $x\in V(F)$, we have $\dist_F(x,\rt(x))\le a_x$ for every outcome of the random choices.
\item The expected total cost of the roots satisfies
\[
\mathbb{E}\Bigl[\sum_{z\in R}a_z\Bigr]
\le \sum_{C\in\textsc{cc}(F)}\min_{v\in C}a_v
+O\Bigl(\sum_{e\in E(F)}\ell_e+\sum_{v\in V(F)}\omega_v\Bigr),
\]
where $\textsc{cc}(F)$ denotes the connected components of $F$ before any edges are removed.
\end{enumerate}
\end{definition}

\paragraph{Input representation for the MPC model.}
As input, we are given $F$ along with its edge weights, vertex weights, and vertex costs.
Here, the hop-diameter of $F$ may be large.
To obtain an $O(1/\sigma)$-round MPC implementation, we additionally assume that a small neighborhood of each vertex is given as input.
More specifically, let $F_x$ be the subgraph of $F$ induced by $\{y\in V(F):\dist_F(x,y)\le a_x\}$.
For each $x\in V(F)$, the vertices and edges of $F_x$, together with their identifiers, weights, and vertex costs, are stored on a single machine. 
In this case, the input size \EMPH{$N$} is 
$\sum_{x\in V(F)}|V(F_x)|$.

\medskip 
In this subsection, we prove the following lemma, which will be used in Section~\ref{sec:euclidean} and Section~\ref{sec:general}. 

\begin{lemma}\label{lem:forestcutting}
Fix a constant $\sigma\in(0,1)$.
Let $F$ be a forest on at most $n$ vertices with positive edge weights $\ell_e$, non-negative vertex weights $\omega_x$, and positive vertex costs $a_x$ such that $a_x\le a_y+\dist_F(x,y)/8$ for any two vertices $x,y$ in the same component of $F$.
Assume that $F$ is stored across machines and that, for each $x\in V(F)$, the records of $F_x$ fit on a single machine with local space $n^\sigma$.
Then there is an $O(1/\sigma)$-round MPC algorithm that computes a low-cost forest partition of $F$ along with the root $\rt(x)$ for every vertex $x\in V(F)$, using $n^\sigma$ local space per machine and $O(N)$ total space.
\end{lemma}

The algorithm consists of two steps.
In the first step, we give orientations to edges and remove some edges so that each resulting component becomes oriented with a unique root. The resulting oriented forest satisfies condition (ii). However, two issues remain: condition (i) does not necessarily hold, and it is unclear if $\rt(x)$ can be computed in $O(1/\sigma)$ rounds. In the second step, we remove some edges while keeping condition (ii) so that,
for each vertex $x\in V(F)$,
either its current root lies in $F_x$, or an edge on the path from $x$ to that root whose endpoint closer to $x$ lies in $F_x$ is removed.
Due to this property, condition (i) is satisfied, and $\rt(x)$ can be computed locally from $F_x$ once the parent and edge-removal information is available.
We describe the first and second steps in Section~\ref{sec:orientation} and Section~\ref{sec:removal}.
Then we provide an MPC implementation in Section~\ref{sec:mpc-forestcutting}.

\subsubsection{Step 1. Giving orientations to edges}
\label{sec:orientation}
The idea is simple:
For each vertex $x\in V(F)$, choose a vertex $v(x)\in V(F_x)$ with the largest weight $\omega_{v(x)}$, breaking ties by a fixed vertex order. If $v(x)\ne x$, orient the first edge on the path from $x$ to $v(x)$ towards $v(x)$.
All choices of $v(x)$ are made in the initial forest $F$.
Then we remove every edge not oriented in this way.
For a technical reason, we also remove the outgoing edge selected by each vertex $x$ with $\omega_x\geq a_x/2$, if such an edge exists. We let $F'$ be the resulting forest, and keep $F$ for the initial forest.

Now we first show that
$F'$ is oriented:
every edge has only one orientation, every component has a single root, and every edge goes towards the root. Then we show that condition (ii) holds.

\begin{lemma}\label{lem:forestcutting-orientation}
Assume $a_x\leq a_y+\dist_F(x,y)/8$
for any two vertices $x,y$ in the same component of $F$.
Then $F'$ is oriented.
\end{lemma}
\begin{proof}
    We claim that every edge $(x,y)$ of $F'$ has only one orientation. 
    Assume that $(x,y)$ is oriented from $x$ to $y$. By construction, 
    $y$ is contained in the path in $F$ between $x$ and $v(x)$. In this case,
    $v(x)\in V(F_y)$. To see this,
    observe that $\dist_F(y,v(x))=\dist_F(x,v(x))-\dist_F(x,y) \leq a_x-\dist_F(x,y) \leq a_y$. 
    The first inequality holds due to $v(x)\in V(F_x)$,
    and the last inequality holds by the assumption that 
    $a_x\leq a_y + \dist_F(x,y)/8$.
    Now assume to the contrary that
    $(x,y)$ is oriented in both directions.
    The argument above implies that 
    $v(x)\in V(F_y)$ and $v(y)\in V(F_x)$.
    By the choice of $v(\cdot)$,
    this implies that $v(x)=v(y)$,
    which is impossible: the unique paths from $x$ and $y$ to the same target vertex $v(x)=v(y)$ cannot both start with $(x,y)$.

    Now observe that every vertex of $F'$ has at most one outgoing edge.
    For a component $C$ of $F'$,
    $|E(C)|=|V(C)|-1$. This implies that
    $C$ has only one vertex without outgoing edges, which is the root of $C$. Moreover, since $F$ has no undirected cycle, every edge of $C$ goes towards its root. This completes the proof of the lemma.     
\end{proof}

\begin{lemma}
  Assume $a_x\leq a_y+\dist_F(x,y)/8$
  for any two vertices $x,y$ in the same component of $F$. 
  Let $R'$ be the set of roots of $F'$. 
  Then we have
\[
\sum_{z\in R'}a_z
\le \sum_{C\in\textsc{cc}(F)}\min_{v\in C}a_v
+O\Bigl(\sum_{e\in E(F)}\ell_e+\sum_{v\in V(F)}\omega_v\Bigr).
\]
\end{lemma}
\begin{proof}
    Let $z$ be a root in $R'$. 
    By construction, there are two cases:
    either $v(z)=z$, or $\omega_z\geq a_z/2$.
    For the latter case, the sum of $a_z$ 
    for all such roots is at most the sum of $2\omega_z$ for all such roots. 
    Therefore, it suffices to focus on the former case. 
    We now analyze an upper bound on $a_z$ for each root $z\in R'$ with $v(z)=z$. For each component $C$ of $F$, choose a vertex $u_C$ with minimum cost $a_{u_C}$ in $C$. For each root $z$ in $C$ with $v(z)=z$, let $\pi_z$ be the undirected path in $F$ starting at $z$ towards $u_C$ and ending at the first vertex $y$ with $\dist_F(z,y)>a_z/8$, or at $u_C$ if no such vertex exists. 
    We use $\ell(\pi_z)$ to denote the total weight of its edges and vertices, excluding its last vertex. If $\dist_F(z,u_C)>a_z/8$, then $a_z<8\ell(\pi_z)$ by construction. Otherwise, $\pi_z$ ends at $u_C$, and the assumption on the vertex costs implies $a_z\leq a_{u_C}+\ell(\pi_z)/8$. \par We now claim that for any two distinct roots $z,z'\in R'$ in the same component of $F$ with $v(z)=z$ and $v(z')=z'$, there is no vertex $y$ satisfying both $\dist_F(z,y)\leq a_z/8$ and $\dist_F(z',y)\leq a_{z'}/8$. Assume to the contrary that such a vertex $y$ exists. Since $y\in V(F_z)\cap V(F_{z'})$, $v(z)=z$, and $v(z')=z'$, we have $\omega_y\leq\min\{\omega_z,\omega_{z'}\}$.
    Thus we have 
    \[
    \begin{aligned}
        \dist_F(z,z')&\leq\dist_F(z,y)+\dist_F(y,z') \\
        &=\dist_F(z,y)+\dist_F(z',y)+\omega_y-\omega_{z'}\\
        &\leq \dist_F(z,y)+\dist_F(z',y) \leq (a_z+a_{z'})/8.
    \end{aligned}
    \]
    The last inequality holds by the assumed bounds on $\dist_F(z,y)$ and $\dist_F(z',y)$. Symmetrically, $\dist_F(z',z)\leq(a_z+a_{z'})/8$. Assume without loss of generality that $z$ is preferred to $z'$ when choosing $v(\cdot)$, that is, either $\omega_z>\omega_{z'}$, or $\omega_z=\omega_{z'}$ and $z$ has a higher priority than $z'$ in the fixed vertex order. By the assumption on the vertex costs, we have $a_z\leq a_{z'}+\dist_F(z,z')/8\leq a_{z'}+(a_z+a_{z'})/64$, which implies $a_z<2a_{z'}$. Therefore, $\dist_F(z',z)<3a_{z'}/8<a_{z'}$, and thus $z\in V(F_{z'})$. This contradicts $v(z')=z'$ by the choice of $v(\cdot)$. This proves the claim. \par By construction, every vertex $y$ of $\pi_z$ other than its last vertex satisfies $\dist_F(z,y)\leq a_z/8$. Thus the claim implies that, after excluding the last vertex of each path, the paths are pairwise vertex-disjoint. Moreover, all paths in a component $C$ go towards $u_C$. If two paths contained the same edge, its endpoint farther from $u_C$ would be a vertex other than the last vertex of both paths, contradicting the claim. Therefore, each edge belongs to at most one path, and the sum of $\ell(\pi_z)$ over all these roots is at most $\sum_{e\in E(F)}\ell_e+\sum_{v\in V(F)}\omega_v$. The claim also implies that, in each component $C$, at most one such root $z$ satisfies $\dist_F(z,u_C)\leq a_z/8$. Hence the term $a_{u_C}=\min_{v\in V(C)}a_v$ is added at most once per component. Summing the bounds on $a_z$ and adding the costs of the roots with $\omega_z\geq a_z/2$ proves the lemma. \end{proof}

\subsubsection{Step 2. Cutting additional edges from the oriented forest}
\label{sec:removal}
Now we have an oriented forest $F'$ satisfying condition (ii). Recall that some edges of the initial $F$ are removed
in the construction of $F'$.
But to simplify notation, we simply redefine $F:=F'$. 
Our goal in this subsection is to remove additional edges from $F$ 
so that condition (i) also holds.
For this, we assign a label $i(x)$ to each vertex $x\in V(F)$, and remove every edge of $F$ whose endpoints have different labels.
In the following, as $F$ is oriented,
the parent $p(x)$ of a non-root vertex $x$ is uniquely defined. We let $\ell_x:=\dist_F(x,p(x))=\ell_{(x,p(x))}+\omega_x$. 
By the construction in Section~\ref{sec:orientation}, we have  $\omega_x<a_x/2$ for every non-root vertex $x$.

\paragraph{Algorithm description.}
We view each edge $(x,p(x))\in E(F)$ as a segment $xp(x)$ of length $\ell_x$. This allows us to refer to points in the interiors of edges. A vertex of $F$ is also considered as a point. 
To allow paths to continue beyond a root, we attach an infinite ray to each root of $F$.
For two points $p$ and $q$ with $q$ on the path from $p$ towards its root or on the attached ray, we extend the notation $\dist_F(p,q)$ to denote the total length of the unique path connecting them, including any partial segments at its ends.

\begin{figure}[t] 
\centering \includegraphics[width=0.7\linewidth]{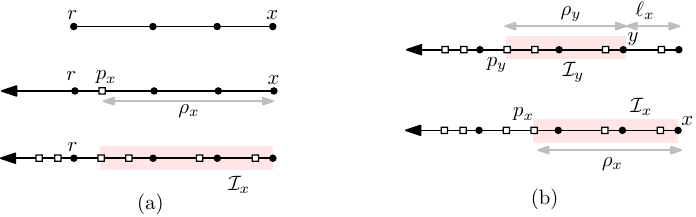} 
\caption{\small Fragments used by the forest cutting algorithm. Black dots denote vertices of $F$, and hollow squares denote marked points $p_z$. Only an ancestor path is shown; other branches are omitted. (a) We attach a ray to the root, mark $p_x$ at distance $\rho_x$ from $x$, and subdivide at all marked points. The shaded fragments form $\mathcal I_x$. (b) For $y=p(x)$, the top and bottom shaded paths consist of $\mathcal I_y$ and $\mathcal I_x$, respectively, and the edge $(x,y)$ has length $\ell_x$.} \label{fig:fragments} 
\end{figure}

For each non-root vertex $x$, set $\rho_x=\max_y\{a_y/2-\dist_F(x,y)\}$, where the maximum is taken over $x$ and its non-root ancestors. For each root $r$, set $\rho_r=a_r/2$. For each vertex $x\in V(F)$, let $p_x$ be the point reached by moving from $x$ towards its root for a distance of exactly $\rho_x$, continuing along the attached ray if necessary.
The point $p_x$ is uniquely defined for each $x\in V(F)$.
We mark the points $p_x$ for all $x\in V(F)$ and subdivide every edge and every attached ray at the marked points. 
Let $\mathcal I$ denote the resulting segments, which we call \EMPH{fragments}, excluding the infinite parts of the infinite rays. Notice that the fragments are pairwise interior-disjoint, and each path from $x$ to $p_x$ is a union of fragments. See Figure~\ref{fig:fragments}(a).
For each fragment $I\in\mathcal I$ of length $\ell(I)$,
we independently draw a \EMPH{random key $\xi_I$} from the exponential distribution with rate $\ell(I)$, that is, $\Pr[\xi_I>t]=e^{-\ell(I)t}$ for $t\geq 0$.

Now we define the label \EMPH{$i(x)$} of a vertex $x\in V(F)$ as follows. We first give a label $i(I)$ to each fragment $I\in\mathcal I$:
if $I$ is contained in an edge of $F$,
it has a unique label, and if it is contained in
an infinite ray attached to a root,
its label is shared by all fragments contained in the same infinite ray. Different rays receive distinct labels that are not assigned to fragments contained in edges of $F$. For each vertex $x\in V(F)$, let $I_x$ be the fragment with the smallest key among those on the path from $x$ to $p_x$, and set $i(x)=i(I_x)$. In particular, each root of $F$ receives the label of its attached ray. Finally, we remove every edge $(x,p(x))\in E(F)$ with $i(x)\ne i(p(x))$ and return the resulting oriented forest \EMPH{$F'$}. 

\paragraph{Analysis.}
We now show that the resulting trees form
a low-cost forest partition of $F$
if $a_x\leq a_y+\dist_F(x,y)/8$ for any two vertices $x,y$ in the same component of $F$, and $\omega_x<a_x/2$ for every non-root vertex $x$. 

\begin{lemma}
For a vertex $x$ of $F$,
we have $a_x/2\leq \rho_x\leq a_x$.
If $p(x)$ is not a root, $|\rho_x-\rho_{p(x)}|\leq\ell_x$.    
\end{lemma}
\begin{proof}
If $x$ is a root, $a_x/2\leq \rho_x\leq a_x$ by definition. Otherwise, $x$ itself is included in the maximum defining $\rho_x$, and hence $\rho_x\geq a_x/2$.
To prove the upper bound, let $y$ be 
a non-root ancestor of $x$, possibly $x$ itself. By the definition of the distance, $\dist_F(y,x)=\dist_F(x,y)+\omega_y-\omega_x\leq\dist_F(x,y)+a_y$, since $\omega_y<a_y/2$ and $\omega_x\geq0$. Applying the assumption on the vertex costs to $(y,x)$ gives \[ a_y\leq a_x+\dist_F(y,x)/8\leq a_x+(\dist_F(x,y)+a_y)/2. \] Rearranging, we obtain $a_y/2-\dist_F(x,y)\leq a_x-\dist_F(x,y)/2\leq a_x$.
Taking the maximum over $y$ proves $\rho_x\leq a_x$. Moreover, if $\rho_x=a_y/2-\dist_F(x,y)$, then $a_x/2\leq \rho_x\leq a_x-\dist_F(x,y)/2$, so $\dist_F(x,y)\leq a_x$ and $y\in V(F_x)$.

Now consider the second part of the lemma.
Let $y=p(x)$ be a non-root vertex. The candidates in the maximum defining $\rho_x$ are $x$, $y$, and the non-root ancestors of $y$. For each candidate other than $x$, the distance from $x$ is $\ell_x$ larger than the distance from $y$. Therefore, \[ \rho_x=\max\{a_x/2,\rho_y-\ell_x\}. \] This immediately gives $\rho_x\geq \rho_y-\ell_x$. On the other hand, the assumption on the vertex costs gives $a_x/2\leq a_y/2+\ell_x\leq \rho_y+\ell_x$. Thus both terms in the maximum are at most $\rho_y+\ell_x$, and hence $\rho_x\leq \rho_y+\ell_x$. Combining the two bounds proves $|\rho_x-\rho_y|\leq\ell_x$.
 \end{proof}
 
 \begin{lemma}\label{lem:forestcutting-max-path-length}
    For a vertex $x\in V(F)$, 
    we have $\dist_F(x,\rt(x))\leq a_x$, where $\rt(x)$ is the root of the tree containing $x$ in $F'$.
\end{lemma}
\begin{proof}
    Let $I_x$ be the fragment with the smallest key among those on the path from $x$ to $p_x$. Then $i(x)=i(I_x)$. 
    If $I_x$ is contained in the infinite ray attached to the root $r$ of the tree in $F$ containing $x$, then $\dist_F(x,r)\leq \rho_x\leq a_x$ by the choice of $I_x$. Since $\dist_F(x,\rt(x))\leq \dist_F(x,r)$, the lemma holds immediately. Now consider the case that 
    an edge $(v,p(v))$ of $F$ contains $I_x$.
    If $i(v)\neq i(x)$, then $\rt(x)$ lies on the path between $x$ and $v$, and $\dist_F(x,\rt(x))\leq\dist_F(x,v)\leq \rho_x\leq a_x$. 
    If this is not the case, $i(v)=i(x)=i(I_x)$, but $i(p(v))\neq i(I_x)$.
    This is because the upward path starting at $p(v)$ does not contain $I_x$, and the label of $I_x$ is unique.
    In this case, the edge $(v,p(v))$ is removed in the construction of $F'$, so $\rt(x)$ again lies on the path between $x$ and $v$.
    Thus the lemma holds in any case.
\end{proof}

\begin{lemma}\label{lem:forestcutting-expected-sum-cost}
    Let $R$ be the set of the roots of $F$,
    and $R'$ be the set of the roots of $F'$.
    Then we have
    \[ \mathbb{E}\Bigl[\sum_{z\in R'\setminus R}a_z\Bigr] \leq \sum_{x\in V(F)\setminus R}O(\ell_x)\] 
    if $a_x\leq a_y+\dist_F(x,y)/8$ for any two vertices $x,y$ in the same component of $F$, and $\omega_x<a_x/2$ for every non-root vertex $x$. 
\end{lemma}
\begin{proof}
    Let $x$ be a non-root vertex of $F$, and let $y=p(x)$. 
    It suffices to show that the probability that $(x,y)$ is cut by the algorithm is $O(\ell_x/a_x)$. Indeed, a non-root vertex $x$ of $F$ becomes a root of $F'$ exactly when $(x,y)$ is cut. By linearity of expectation, we have \[ \mathbb E\Bigl[\sum_{z\in R'\setminus R}a_z\Bigr] =\sum_{(x,y)\in E(F)}a_x\Pr[i(x)\neq i(y)] \leq \sum_{x\in V(F)\setminus R} O(\ell_x). \]

    Now we focus on analyzing the probability that $(x,y)$ is cut. 
    Let $\mathcal I_x$ be the set of fragments contained in the path from $x$ to $p_x$.
    Similarly, let $\mathcal I_y$ be the set of fragments contained in the path from $y$ to $p_y$. See Figure~\ref{fig:fragments}(b). First suppose that $y\notin R$. The event that $(x,y)$ is cut by the algorithm happens only when 
    the fragment of $\mathcal I_x\cup\mathcal I_y$ with the smallest key is contained in $\mathcal I_x\oplus\mathcal I_y$,
    where
    $\mathcal I_x\oplus \mathcal I_y=(\mathcal I_x\setminus \mathcal I_y)\cup(\mathcal I_y\setminus \mathcal I_x)$. Then we have \[ \Pr[i(x)\neq i(y)] \leq
    \Pr\Bigl[\operatorname*{arg\,min}_{I\in\mathcal I_x\cup \mathcal I_y} \xi_I \in \mathcal I_x\oplus\mathcal I_y\Bigr]
    = \frac{\sum_{I\in\mathcal I_x\oplus\mathcal I_y}\ell(I)}{\sum_{I\in\mathcal I_x\cup\mathcal I_y}\ell(I)} \leq O(\ell_x/a_x). \] 
    Here, the equality holds
    by a standard property of independent exponential random variables, 
    and the last inequality holds since      
    the total length of the fragments in $\mathcal I_x\oplus\mathcal I_y$ is at most $\ell_x+|\ell_x+\rho_y-\rho_x|\leq 3\ell_x$
    by the bound $|\rho_x-\rho_y|\leq \ell_x$ proved above, and 
    the total length of the fragments in $\mathcal I_x\cup\mathcal I_y$ is at least $\rho_x\geq a_x/2$.
     
     Now suppose that $y\in R$. The label $i(y)$ is shared by all fragments on the ray attached to $y$. Thus $i(x)\neq i(y)$ exactly when $I_x$ lies on the edge $(x,y)$. The portion of the path from $x$ to $p_x$ on this edge has length $\min\{\rho_x,\ell_x\}$, and hence $\Pr[i(x)\neq i(y)]=\min\{\rho_x,\ell_x\}/\rho_x\leq\ell_x/\rho_x\leq 2\ell_x/a_x$.
\end{proof}

\subsubsection{MPC implementation}
\label{sec:mpc-forestcutting}
We now provide an MPC construction of a low-cost forest partition of $F$ using the primitives in Section~\ref{sec:prelim}.
As input, we are given $F$ along with its edge weights $\ell_e$, vertex weights $\omega_x$, vertex costs $a_x$, and neighborhoods $F_x$.
For each $x\in V(F)$, the vertices and edges of $F_x$ are stored on a single machine, and the input size is $N=\sum_{x\in V(F)}|V(F_x)|$.

\paragraph{First step.}
In the first step, we choose $v(x)\in V(F_x)$ and the parent edge, if any, of each vertex $x$ locally, since $F_x$ contains the path from $x$ to $v(x)$ and is stored in a single machine.
Using \textsc{Aggregation} on edge identifiers, we identify the edges that are not chosen by any vertex.
Removing the outgoing edge of a vertex $x$ with $\omega_x\geq a_x/2$ can be done locally.
Using \textsc{Broadcasting}, we send the parent and parent-edge length of each vertex, or its root status, to every $F_x$ containing that vertex.
Therefore, the first step can be implemented in $O(1/\sigma)$ MPC rounds.

\paragraph{Second step.}
In the second step, we compute $\ell_x$ for non-root vertices and $\rho_x$ and $p_x$ for all vertices locally using the parent information received in the first step.
The proof above shows that the maximum defining $\rho_x$ can be computed using only $x$ and its non-root ancestors in $F_x$.
Using \textsc{Sorting}, we sort the marked points $p_x$ along each edge and attached ray and obtain $O(n)$ fragments, keeping only the part up to the last marked point on each ray.
Then we choose a random key $\xi_I$ and the label $i(I)$ of each fragment $I$ locally.\footnote{We assume that $\xi_I$ fits within a single word,
which is also assumed in~\cite{azarmehr2025massively}.}
Using the machine ID and a local index, we assign distinct labels to the fragments contained in edges of $F$. All fragments on the ray attached to a root share the label of that root, using labels distinct from those of fragments in edges.

Now we assign the label $i(x)$ for each vertex $x$ as follows.
Note that the number of fragments contained in the $x$-$p_x$ path can be large although the number of edges contained in the path is $O(|V(F_x)|)$.
To resolve this issue, for each edge $e$ of $F$, we compute the minimum key among its fragments together with the label of the corresponding fragment using \textsc{Aggregation}.
Using \textsc{Broadcasting}, we send this pair to every $F_x$ containing $e$.
Thus each $F_x$ receives $O(|V(F_x)|)$ pairs.
Since the total size of the neighborhoods stored on each machine is $O(n^\sigma)$, each machine receives $O(n^\sigma)$ pairs, and the total number of pairs is $O(N)$.

For each $p_x$, we also compute the minimum key and its corresponding label on the part of the last edge or ray on the $x$-$p_x$ path ending at $p_x$.
This part starts at the child endpoint of the edge or at the root of the ray.
We obtain these $O(n)$ pairs using \textsc{PrefixSum} along each edge and ray, and send the pair for $p_x$ to the machine containing $F_x$ using \textsc{Broadcasting}.
Then each vertex $x$ obtains $i(x)$ locally from the pairs for the complete edges on its path and the part ending at $p_x$.
Finally, we use \textsc{Broadcasting} to collect the endpoint labels of each edge, compare them to determine whether the edge is removed, and send this information to every $F_x$ containing its child endpoint.
Then each vertex $x$ can find its root locally in $F_x$ by Lemma~\ref{lem:forestcutting-max-path-length}.
We use a constant number of primitive operations on $O(N)$ records, and thus spend $O(1/\sigma)$ MPC rounds with $O(N)$ total space.
This completes the proof of Lemma~\ref{lem:forestcutting} together with all arguments in Section~\ref{sec:orientation} and Section~\ref{sec:removal}.

\section{$(1+\eps)$-approximate MST in general metrics}
\label{sec:general}
In this section, we present a fully scalable MPC algorithm for computing a $(1+\eps)$-approximate MST of 
a general metric $(X,\dist)$ with $n$ vertices.
For any fixed constants $\sigma,\eps\in(0,1)$, the algorithm uses $O((1+\log(1/(\sigma\eps)))/\sigma^2)$ rounds, $n^\sigma$ local space per machine, and $\widetilde{O}(n^2)$ total space. Each primitive in Section~\ref{sec:prelim} takes $O(1/\sigma)$ rounds on a polynomial number of records~\cite{GoodrichSitchinavaZhang2011Sorting}. 
We may consider the metric $(X,\dist)$ as a 
complete undirected graph whose edge weights are the distances of a metric on its vertex set,
and call it the \EMPH{metric graph} of $(X,\dist)$.
Let $G_0$ be the input metric graph of $(X,\dist)$. 

We construct a spanning tree of \(G_0\) in a bottom-up fashion. 
At each level, we partition the current set \(Q\) of vertices into clusters and construct a spanning tree \(T_C\) on each cluster \(C\). 
Then we contract each cluster \(C_v\) into a single vertex \(v\) and proceed with the contracted graph. 
Thus each current vertex represents a connected component of the forest constructed so far in \(G_0\). Throughout the construction, for each current edge, we store an edge of \(G_0\) of minimum weight between the two represented components and use its weight as the current edge weight. Before contracting each cluster \(C\), we add the stored edge of \(G_0\) corresponding to each edge of \(T_C\) to the final spanning tree.
Since distances between contracted vertices need not satisfy the triangle inequality, we assign each vertex \(v\) a weight \(\omega(v)\) equal to the total edge weight of the tree maintained within its component. With these vertex weights, we have \(\dist(u,v)\leq\dist(u,z)+\omega(z)+\dist(z,v)\), since the tree within the component represented by \(z\) connects any two of its points at cost at most \(\omega(z)\).

\subsection{Overall algorithm}
To formulate this idea, we introduce the notion of a \emph{light contraction}.
Let $G$ be a complete graph with edge weight $\dist(\cdot,\cdot)$
and nonnegative vertex weight $\omega(\cdot)$.
We use \EMPH{$\opt(G)$} to denote the minimum total edge weight and vertex weight of a spanning tree of $G$.
For a partition $\mathcal C$ of $V(G)$,
the contracted graph \EMPH{$G/\mathcal C$} is defined as
the complete graph where a vertex $v$ corresponds to a partition class $C_v$ of 
$\mathcal C$.
We set the edge weight \EMPH{$\dist_\mathcal C (u,v):=\dist(C_u,C_v)$}. 
Consider a spanning tree $T_C$ of a partition class $C$ of $\mathcal C$.
Its weight $\weight(T_C)$ is defined as 
the total edge weight and vertex weight of $T_C$.
For each vertex $v\in V(G/\mathcal C)$,
we set the vertex weight \EMPH{$\omega_\mathcal C(v):=\weight(T_{C_v})$}.
When the context is clear,
we omit the subscript $\mathcal C$ from $\dist_\mathcal C(\cdot,\cdot)$ and $\omega_\mathcal C(\cdot)$.

\begin{definition}[Light contraction]
\label{def:light-contraction}
    Let $b\geq 2$ and $\eps\in(0,1)$, and let $G$ be an $m$-vertex complete graph with edge-weight function $\dist(\cdot,\cdot)$ and nonnegative vertex-weight function $\omega(\cdot)$.
    Consider a partition $\mathcal C$ of $V(G)$ and spanning trees $\{T_C\}_{C\in\mathcal C}$.
    Then we say $G/\mathcal C$ is 
    a \EMPH{$(b,\eps)$-light contraction} of $G$ if the following conditions hold:
    \begin{itemize}
        \item[(i)] The number of vertices of $G/\mathcal C$ is $O(m/b)+1$.
        \item[(ii)] The MST cost of the contracted graph satisfies
            $\opt(G/\mathcal C)
            \leq (1+\eps)\cdot \opt(G)$.
    \end{itemize}
\end{definition}

A main technical contribution of this section is an $O((1+\log(1/\eps))/\sigma)$-round randomized MPC construction of a light contraction $G/\mathcal C$,
along with a partition $\mathcal C$ of $V(G)$ and spanning trees $\{T_C\}_{C\in\mathcal C}$, which is stated in the following lemma.

\begin{lemma}
\label{lem:light-contraction}
    Fix two parameters $\eps,\sigma\in(0,1)$. Let $n$ and $m$ be two integers with $m\leq n$, and let $b=n^{\sigma/10}$.  
    Given an $m$-vertex complete graph $G$ with edge weight $\dist(\cdot,\cdot)$
    and vertex weights $\omega(\cdot)$ satisfying
    $\dist(u,v)\leq \dist(u,z)+\omega(z)+\dist(z,v)$ for any three vertices $u,v,z\in V(G)$,  
    there is a randomized MPC algorithm that
    computes a partition $\mathcal C$ and spanning trees
    $\{T_C\}_{C\in\mathcal C}$ such that $G/\mathcal C$ is a
    $(b,\eps)$-light contraction with probability at least
    $1-n^{-\Omega(1)}$.
    The algorithm uses $O(\log(1/\eps)/\sigma)$ rounds, $n^\sigma$ local space
    per machine, and $\widetilde{O}(n^2)$ total space.
\end{lemma}

We first prove Theorem~\ref{thm:general-metric-mst} using Lemma~\ref{lem:light-contraction}. We then describe the construction of a light contraction and its MPC implementation in Section~\ref{sec:general-contraction}.

\begin{proof}[Proof of Theorem~\ref{thm:general-metric-mst} assuming Lemma~\ref{lem:light-contraction}]
Let $L=\lceil 20/\sigma\rceil$, and keep $b=n^{\sigma/10}$ fixed throughout the algorithm.
Initially, the weight $\omega(\cdot)$ of every vertex of $G_0$ is zero, and the weight of each edge $uv$ is $\dist(u,v)$.
Starting with $G=G_0$, repeatedly compute a $(b,\eps/(2L))$-light contraction using Lemma~\ref{lem:light-contraction} while $|V(G)|>n^{\sigma/3}$.
For each cluster $C$, keep the stored original edges corresponding to the edges of $T_C$ as part of the output and contract $C$.
We make at most $L$ calls and report failure if more than $n^{\sigma/3}$ vertices remain.

Each tree $T_C$ joins distinct current components, so the selected original edges remain a forest.
The updated vertex weights are the weights of the maintained internal trees, so the inequality required by the lemma continues to hold.
On every successful call, the number of vertices becomes $O(m/b)+1$; while $m>n^{\sigma/3}$, this is at most $m/n^{\sigma/20}$ for sufficiently large $n$.
Thus $L$ successful calls suffice.
The remaining complete graph occupies $O(n^{2\sigma/3})$ words and fits on one machine.
Compute its exact MST and add the corresponding original edges to the output.
The resulting spanning tree $T$ has $w(T)=\opt(G)$ for the final graph $G$.
If all calls succeed, then
\[
    w(T)\leq \left(1+\frac{\eps}{2L}\right)^L\opt(G_0)
    \leq e^{\eps/2}\opt(G_0)
    \leq (1+\eps)\opt(G_0).
\]

Sorting and aggregation update the vertex weights and the minimum edges between the new components, together with their stored original edges, in $O(1/\sigma)$ rounds using $O(n^2)$ total space.
With fresh randomness in each call, a union bound over at most $L$ calls gives success probability $1-n^{-\Omega(1)}$.
Since $L=O(1/\sigma)$ and each call uses error $\eps/(2L)$, the total number of rounds is $O((L/\sigma)(1+\log(2L/\eps)))=O((1+\log(1/(\sigma\eps)))/\sigma^2)$, and the total space is $\widetilde O(n^2)$.
\end{proof}

\subsection{Proof of Lemma~\ref{lem:light-contraction}: Computing a light contraction}
\label{sec:general-contraction}
In this subsection, we prove Lemma~\ref{lem:light-contraction}: compute a $(b,\eps)$-light contraction of
an $m$-vertex complete graph $G$ with edge weight $\dist(\cdot,\cdot)$
and vertex weights $\omega(\cdot)$.
Throughout this subsection, we assume the following weighted triangle inequality:
\begin{equation}
\label{eqn:general-weighted-triangle}
\dist(u,v)\leq\dist(u,z)+\omega(z)+\dist(z,v)
\qquad\text{for all }u,v,z\in V(G).
\end{equation}
Here, $\eps,\sigma \in(0,1)$, $n\geq m$, and \EMPH{$b$}$ =n^{\sigma/10}$.
    In particular, $(b\log n)^2$ is asymptotically smaller than the local space complexity. 
   We use $O(n^\sigma)$ local space and $\widetilde O(n^2)$ total space. 
For a vertex $v\in V(G)$ and a radius $r$, we use \EMPH{$B_G(v,r)$}$:=\{u\in V(G)\mid \omega(v)+\dist(v,u)\leq r\}$.
 
\medskip 
We start by computing the smallest value \EMPH{$r_v$} 
with $|B_G(v,r_v)|=\min\{m,\lceil c b\log n\rceil\}$ for every vertex $v\in V(G)$, and we let \EMPH{$S_v$}$:= B_G(v,r_v)$, where $c>0$ is a sufficiently large constant.
Since we use $\Theta(n^2)$ total space, we can compute $r_v$ and $S_v$ in $O(1/\sigma)$ rounds for every vertex $v\in V(G)$ using \textsc{Sorting}.
Then we apply \textsc{TruncatedPrim}$(v,S_v,r_v/2,\omega)$ for every vertex $v\in V(G)$. Here, $\omega$ denotes the vertex weights $\{\omega(v)\}_{v\in V(G)}$. We place all vertices of $S_v$ and all edges in $S_v\times S_v$ into a single machine for each vertex $v\in V(G)$ in $O(1/\sigma)$ MPC rounds in total.
Then we run \textsc{TruncatedPrim} locally without any communication.
Let $F$ be the union of the edge sets returned by the calls of \textsc{TruncatedPrim}.
Then the edges and vertices of $F$ are spread over the machines, and some edges are created multiple times by different calls.
We remove duplicate copies in $O(1/\sigma)$ MPC rounds using \textsc{Sorting}.
By Lemma~\ref{lem:truncated-prim}, $F$ is contained in an MST of $G$.
Therefore, condition (ii) holds immediately by taking $\mathcal C$ as the vertex sets of the connected components of $(V(G),E(F))$ and each $T_C$ as the corresponding component tree. Indeed, we have $\opt(G/\mathcal C)=\opt(G)$.

However, we still have two issues: we need to identify the component of $F$ containing each vertex, and
condition (i) does not necessarily hold.
Since each tree of $F$ might be large, it is unclear if we can do this in $O(1/\sigma)$ MPC rounds.
To resolve the first issue, we apply \textsc{ForestCutting} described in Section~\ref{sec:subroutine},
and to resolve the second issue, we merge several components so that we have $O(m/b)+1$ components at the end.

\subsubsection{Forest cutting} 
We first apply \textsc{ForestCutting} with \EMPH{$a_v:=r_v/8$} and \EMPH{$\ell_{uv}:=\dist(u,v)$} for every vertex $v\in V(F)$ and edge $(u,v)\in E(F)$. The subroutine \textsc{ForestCutting} requires $F_x$ to be stored on a single machine for each $x\in V(F)$, where 
\EMPH{$F_x$} is the subgraph of $F$ induced by $\{y\in V(F) : \dist_F(x,y)\leq r_x/8\}$.
By construction, $F_x$ has $O(b\log n)$ vertices, and thus we can place all edges and vertices of $F_x$ on a single machine in $O(1/\sigma)$ MPC rounds. 
The total space of $F_x$ for all vertices $x\in V(F)$ is within the total space complexity of $O(n^2)$.

We first show that this satisfies the assumption of Lemma~\ref{lem:forestcutting} in the following lemma.
In Section~\ref{sec:subroutine}, recall that we let $\dist_F(x,y)$ be
the total weight of the edges and vertices on their unique path, including $x$ but excluding $y$, for any two distinct vertices $x$ and $y$ in the same component of $F$. We set $\dist_F(x,x)=0$.

\begin{lemma}
    For any edge $(u,v)\in E(F)$, $r_u\leq r_v+\omega(u)+\dist(u,v)$.
    Therefore, $a_x\leq a_y+\dist_F(x,y)/8$ for any two vertices $x,y\in V(F)$ in the same component of $F$.
\end{lemma}
\begin{proof}
Recall that $r_v$ is the smallest radius such that $|B_G(v,r_v)|=\min\{m,\lceil c b\log n\rceil\}$, and $S_v=B_G(v,r_v)$. For every $z\in S_v$, Equation~(\ref{eqn:general-weighted-triangle}) gives
\[
\omega(u)+\dist(u,z)
\leq \omega(u)+\dist(u,v)+\omega(v)+\dist(v,z)
\leq \omega(u)+\dist(u,v)+r_v.
\]
Here, the second inequality follows from the fact that
$z\in S_v$. Thus $S_v\subseteq B_G(u,\omega(u)+\dist(u,v)+r_v)$, so this ball contains at least $\min\{m,\lceil c b\log n\rceil\}$ vertices. By the definition of $r_u$, we obtain
$r_u\leq \omega(u)+\dist(u,v)+r_v$, which proves the first part of the lemma. 

This also immediately gives $a_u\leq a_v+\dist_F(u,v)/8$ by the definition of $\dist_F(\cdot,\cdot)$ and $a_v$ for an edge $(u,v)$ of $F$. 
Summing these inequalities along the $x$-$y$ path in $F$ gives
$a_x\leq a_y+\dist_F(x,y)/8$ for any two vertices $x,y$ in the same component of $F$.
\end{proof}

Thus we can apply \textsc{ForestCutting} on $F$ in $O(1/\sigma)$ MPC rounds. Then we have a low-cost forest partition $\mathcal F$ of $F$ consisting of oriented trees
such that 
\begin{equation}
\label{eqn:root-cost-general}
\mathbb{E}\Bigl[\sum_{z\in R}r_z/8\Bigr]
\le \sum_{C\in\textsc{cc}(F)}\min_{v\in C} r_v/8
+O\Bigl(\sum_{(u,v)\in E(F)}\dist(u,v)+\sum_{v\in V(F)}\omega(v)\Bigr),
\end{equation}
where $\textsc{cc}(F)$ denotes the connected components of $F$ before any edges are removed and $R$ denotes the roots of the trees of $\mathcal F$, by Definition~\ref{def:low-cost-forest-partition} and Lemma~\ref{lem:forestcutting}.
Moreover, each vertex $v$ has the identifier of its root $\rt(v)$ of the tree containing $v$ in $\mathcal F$, which allows us to identify the component of $\mathcal F$ containing $v$. 
Also, condition (ii) still holds, since the edges retained by \textsc{ForestCutting} are contained in an MST of $G$.

Before discussing the second step, we prove the following lemma, which will be used in proving that condition (ii) still holds after the second step.
\begin{lemma}\label{lem:general-connecting-cost}
    $\mathbb{E}\Bigl[\sum_{z\in R}r_z\Bigr]=O(\opt(G))$,
    where $R$ denotes the set of the roots of the trees of $\mathcal F$.
\end{lemma}
\begin{proof}
It suffices to show that the right-hand side of Equation~(\ref{eqn:root-cost-general}) is $O(\opt(G))$. The second term of the right-hand side of the equation is $O(\opt(G))$ since  $F$ is contained in the MST of $G$. Thus we focus on proving 
$\sum_{C\in\textsc{cc}(F)}\min_{v\in C}r_v=O(\opt(G))$.

If $F$ has a single component, 
$\omega(v)+\dist(v,x)\leq\weight(F)$ for all vertices $v,x\in V(F)$
by applying Equation~(\ref{eqn:general-weighted-triangle}) along the $v$-$x$ path in $F$.
This implies that $B_G(v, \weight(F))$ contains all vertices of $F$.
By the choice of $r_v$, we have $r_v\leq\weight(F)=\opt(G)$, which
is the desired bound.

Otherwise, let $T$ be an MST of $G$ containing $F$.
For each component $C$ of $F$, choose an edge $(u,v)\in E(T)$ with $u\in C$ and $v\notin C$.
If $v\notin S_u$, $r_u<\omega(u)+\dist(u,v)\leq \weight(F[C])+\dist(u,v)$.
Otherwise, consider the call \textsc{TruncatedPrim}$(u,S_u,r_u/2,\omega)$.
The tree visited by this call is contained in $F[C]$.
There are two cases when the subroutine terminates: either the total weight of the edges and vertices visited by the call is at least $r_u/2$, or the next edge has length at least $r_u/2$.
In the first case, $r_u\leq2\cdot\weight(F[C])$.
In the second case, $r_u\leq2\cdot\dist(u,v)$, since $u$ is visited and $v\notin C$.
Thus, in all three cases,
\[
\min_{x\in C}r_x\leq r_u\leq2\cdot\weight(F[C])+2\cdot \dist(u,v).
\]
Each edge $(u,v)$ of $T$ is chosen by at most two components of $F$, one containing $u$ and one containing $v$.  Therefore,
\[
\sum_{C\in\textsc{cc}(F)}\min_{x\in C}r_x
\leq\sum_{C\in\textsc{cc}(F)}2\cdot\weight(F[C])
+\sum_{(u,v)\in E(T)}4\cdot\dist(u,v)
\leq6\cdot\opt(G),
\]
which proves the lemma.
\end{proof}

\subsubsection{Forest merging}
Now we merge components of $\mathcal F$ by adding edges of $G$, while keeping the additive error induced by this procedure at most $\eps\cdot\opt(G)$.
We may use a similar approach to that used in the Euclidean case, which we call \textsc{Merging-via-Sampling}:
sample each component with probability proportional to its size, and for a component which is not sampled, connect it to the closest sampled component.\footnote{Unlike in the Euclidean case, we can find the closest sampled component for each unsampled component in $O(1/\sigma)$ MPC rounds in total as we have $\Theta(n^2)$ total space.}
Here, the size of a component is the \emph{number} of vertices of $G$ it contains.
As in the Euclidean case analyzed in Section~\ref{sec:euclidean}, we can show that
the additive error induced by this procedure is at most $\sum_{z\in R}r_z$, which is $O(\opt(G))$ in expectation by Lemma~\ref{lem:general-connecting-cost}.
This is acceptable in Section~\ref{sec:euclidean}, but our desired approximation ratio is $(1+\eps)$ here.

To reduce this expected error, we first merge several components using edges of an MST $T$ of $G$ before applying \textsc{Merging-via-Sampling}.
This reduces the expected sum of $r_z$ of the roots of the resulting trees to $\eps'\cdot \opt(G)$ for a sufficiently small constant $\eps'\leq \eps$. 
For this, we use the cut property: 
for each component $C$ of $\mathcal F$, the minimum-weight edge of $G$ leaving $C$ is contained in $T$. Therefore, we may merge the two components containing its endpoints without introducing any error. However, these edges may connect many components in a long chain, so it is unclear how to identify the resulting component containing each vertex in $O(1/\sigma)$ MPC rounds.
To resolve this, we use the standard technique of leader compression
and call the resulting procedure \textsc{Merging-via-Leader-Compression}.

\paragraph{Merging via leader compression.}
We repeat the following procedure, starting with the components of $\mathcal F$ and using their roots as representatives. In each iteration, the expected sum of $r_z$ for all roots $z$ decreases by a constant fraction, counting this sum as zero once only one component remains.

In each iteration, 
for each current component $C$, we
pick an edge $e_C=(u,v)$ with $u\in C$ and $v\notin C$
with minimum $\dist(u,v)$ in $O(1/\sigma)$ MPC rounds in total for all components. 
Then each component becomes a leader independently with probability $1/2$.
For each nonleader component $C$,
let $e_C=(u,v)$ with $u\in C$ and $v\notin C$. 
If the component containing $v$ is a leader, we add $e_C$ and merge $C$ with that component.
We do this for all components in $O(1/\sigma)$ MPC rounds.
Each component formed by a merge consists of one leader component, say $C_\text{lead}$, and nonleader components. A nonleader component whose chosen edge does not lead to a leader remains unchanged.
Every edge used to merge these components has one endpoint in $C_\text{lead}$.  
For each new component, we use the root of $C_\text{lead}$ as the root of the merged component.
Thus, using \textsc{Sorting} and \textsc{Broadcasting},
we can pass the identifier of the new root to every vertex of its component in $O(1/\sigma)$ MPC rounds.

We repeat this procedure for a sufficiently large number $t=O(1+\log(1/\eps'))$ of iterations, stopping as soon as only one component remains. 
Each iteration takes $O(1/\sigma)$ MPC rounds, so the entire procedure takes $O((1+\log(1/\eps'))/\sigma)$ rounds.
    
\begin{lemma}
After applying \textsc{Merging-via-Leader-Compression},
$\mathbb{E}\Bigl[\sum_{z\in R'}r_z\Bigr]\leq(\eps'/2)\cdot\opt(G)$ 
if there are at least two components,
where $R'$ denotes the set of roots of the resulting forest.
\end{lemma}
\begin{proof}
Let $\Phi_i$ be the sum of $r_z$ over the roots after $i$ iterations.
Initially, $\mathbb{E}[\Phi_0]=O(\opt(G))$ by Lemma~\ref{lem:general-connecting-cost}.
Fix an iteration where the current forest has at least two components.
For each component $C$, the probability that $C$ is a nonleader and the component reached by $e_C$ is a leader is $1/4$.
If this event happens, $C$ is merged into the leader component. 
Thus with probability $1/4$, each $r_z$ is removed from $\Phi_i$. 
Therefore,
$\mathbb{E}[\Phi_{i+1}\mid\text{the current forest}]
\leq(3/4)\cdot \Phi_i$.
Thus, after a sufficiently large number $t=O(1+\log(1/\eps'))$ of iterations,
$\mathbb{E}[\Phi_t]
\leq (3/4)^t\cdot\mathbb{E}[\Phi_0]
\leq(\eps'/2)\cdot\opt(G)$, 
which proves the lemma.
\end{proof}

\paragraph{Merging via sampling.}
Now we merge components so that the number of resulting components becomes $O(m/b)+1$. This is the only step that can increase the MST cost.
For each component $C$ returned by \textsc{Merging-via-Leader-Compression}, we sample it independently with probability
$p_C=\min\{1,|C|/b\}$,
where $|C|$ denotes the number of vertices of $G$ in $C$.
The expected number of initially sampled components is at most $m/b$, since $\sum_C|C|=m$. If no component is sampled, we choose an arbitrary component and treat it as sampled.
Then for each non-sampled component $C$,
we find its closest sampled component $\textsc{nn}(C)$,
that is, we find the edge $(u,v)$ with minimum $\dist(u,v)$ such that $u\in C$ and the component $\textsc{nn}(C)$ containing $v$ is sampled, and then
connect $C$ with $\textsc{nn}(C)$ using $(u,v)$.
Since we have $\Theta(n^2)$ total space,
this can be done in $O(1/\sigma)$ MPC rounds for all  non-sampled components.

\begin{lemma}\label{lem:general-sampling-cost}
Let $A$ be the set of edges added by \textsc{Merging-via-Sampling}.
Then
\[
\mathbb{E}\Bigl[\sum_{(u,v)\in A}\dist(u,v)\Bigr]
\leq\eps'\cdot\opt(G).
\]
\end{lemma}
\begin{proof}
Let $\mathcal E$ be the event that, for every non-sampled component $C$ with root $z$, some sampled component intersects $S_z$.
Condition on the forest after leader compression.
If $m\leq\lceil c b\log n\rceil$, then $S_z=V(G)$ for every root $z$, so $\mathcal E$ holds since at least one component is sampled.
Otherwise, for each root $z$, the probability that no initially sampled component intersects $S_z$ is at most
\[
\prod(1-p_C)
\leq \exp(-|S_z|/b)
\leq \exp(-c\log n),
\]
where the product is over the current components $C$ intersecting $S_z$, whose sizes sum to at least $|S_z|$.
A union bound over at most $m\leq n$ roots gives $\Pr[\mathcal E^c]\leq\eps'/(2n)$ for a sufficiently large constant $c$ depending on $\eps'$, where $\mathcal E^c$ denotes the event that $\mathcal E$ does not occur.

Fix an outcome in $\mathcal E$ and a non-sampled component $C$ with root $z$.
There is a sampled component $C'$ containing a vertex $x\in S_z$.
Since $z\in C$, the edge $(z,x)$ is a candidate for connecting $C$ to a sampled component.
Thus, the edge $(u,v)$ added for $C$ satisfies
\[
\dist(u,v)\leq\dist(z,x)\leq r_z-\omega(z)\leq r_z,
\]
where the second inequality follows from the definition of $S_z$.

For every pair $u,v\in V(G)$, applying Equation~(\ref{eqn:general-weighted-triangle}) along the $u$-$v$ path in an MST of $G$ gives $\dist(u,v)\leq\opt(G)$.
Since at most $m-1\leq n$ edges are added, the total added weight is at most $n\cdot\opt(G)$ even when $\mathcal E$ does not hold.
Therefore,
\[
\mathbb{E}\Bigl[\sum_{(u,v)\in A}\dist(u,v)\Bigr]
\leq\mathbb{E}[\Phi_t]+n\cdot\Pr[\mathcal E^c]\cdot\opt(G)
\leq\eps'\cdot\opt(G).\qedhere
\]
\end{proof}

Let $\mathcal C$ be the final partition of $V(F)$.
Note that each component $C$ of $\mathcal C$ has a spanning tree $T_C$, and each vertex $v\in C$ is associated with its root $\rt(v)$ in $T_C$.
Therefore, we can compute $G/\mathcal C$ in $O(1/\sigma)$ MPC rounds using $O(n^2)$ total space.
We have shown that the expected number of components of $\mathcal C$ is $O(m/b)$.
By Lemma~\ref{lem:general-sampling-cost},
the expected value of $\opt(G/\mathcal C)$ is at most 
$(1+\eps)\cdot \opt(G)$.
By combining these results, we now prove Lemma~\ref{lem:light-contraction}.

\begin{proof}[Proof of Lemma~\ref{lem:light-contraction}]
We apply the entire construction $\Theta(\log n)$ times independently in parallel with $\eps'=\eps/4$. 
For each execution, let $K$ be the number of components in $\mathcal C$, and let $W$ be the total weight of the edges added by \textsc{Merging-via-Sampling}.
We already showed that $\mathbb E[K]\leq O(m/b)$ and $\mathbb E[W]\leq(\eps/4)\cdot\opt(G)$.
By Markov's inequality and a union bound, the two bounds
$K\leq O(m/b)$ and $W\leq\eps\cdot\opt(G)$ 
hold simultaneously with probability at least $1/2$.
Furthermore, we already showed that for every execution,
$\opt(G/\mathcal C)\leq\opt(G)+W$.

Among the $\Theta(\log n)$ executions with $K\leq O(m/b)$, we return the one with minimum $W$, and report failure if there is no such execution.
With probability at least $1-n^{-\Omega(1)}$, some execution satisfies both bounds above.
The selected execution then also satisfies both bounds, and hence gives a $(b,\eps)$-light contraction. The executions are performed in parallel, and aggregation computes their values of $K$ and $W$ and selects an execution in $O(1/\sigma)$ rounds. With $\eps'=\eps/4$, the total number of rounds is therefore $O((1+\log(1/\eps))/\sigma)$, and the total space is $\widetilde O(n^2)$.
\end{proof}

\section{Constant approximate MST in Euclidean and doubling spaces}
\label{sec:euclidean}
In this section, we present fully scalable MPC algorithms for computing an $O(1)$-approximate MST of a set $P$ of $n$ points in the Euclidean space of dimension $d$ and a doubling space of doubling dimension $\ddim$.
For any fixed constant $\sigma\in(0,1)$, our algorithms use $O(1/\sigma^2)$ and $O(\ddim/\sigma^4)$ rounds in Euclidean spaces and doubling metrics, respectively.
The corresponding total space bounds are $O(nd+n^{1+\sigma})$ and $\widetilde O(n^{1+\sigma})\cdot 2^{O(\ddim)}$, respectively.
For a point set $Q\subseteq P$, let \EMPH{$\opt(Q)$} denote the weight of an MST of $Q$.
Let $G$ be a graph such that a vertex of $G$ corresponds to a point in $P$.
We use \EMPH{$\omega(G)$} to denote the sum of its edge weights, where the weight of an edge is the distance between its endpoints.
To avoid heavy notation, we also use $\omega(G)$ instead of $\weight(G)$ for a graph $G$.

As for general metrics, our algorithm constructs a spanning tree in a bottom-up fashion, starting with the input point set $P$.
At each level, we partition the current point set $Q$ into clusters and construct a spanning tree $T_C$ on each cluster $C$.
In the general metric case, we contract each cluster into a single vertex and proceed with the contracted graph.
We instead select one representative point per cluster for the next level.
We repeat this until the remaining representatives fit on a single machine.
At this level, we can compute the MST of the representatives on a single machine. 
Then we return the tree edges computed at all levels as the output.

Here, for $d$-dimensional Euclidean spaces, we assume $d=O(\log n)$; if this is not the case, we apply the Johnson-Lindenstrauss reduction~\cite{JohnsonLindenstrauss1984Extensions}, which can be implemented in constant MPC rounds~\cite{jayaram2024massively} in the fully scalable MPC model.
Here, all pairwise distances are preserved within a constant factor with high probability.

\subsection{Overall algorithm}
To formulate this idea, we require that the number of representatives decreases by a polynomial factor at each level.
To achieve an approximation ratio of $O(1)$, we construct spanning trees for all clusters at every level with expected total weight $O(\opt(Q))$, where $Q$ is the current representative set.
For this, we introduce the notion of a light partition.

\begin{definition}[Light partition]
    Let $b\geq 2$ be an integer, and let $\lambda \geq 1$ be a parameter.
    Let $(X,\dist)$ be a metric space.
    For a set $P$ of $n$ points in $(X,\dist)$, a \EMPH{$(b,\lambda)$-randomized light partition} is a partition $\mathcal C$ of $P$, together with a spanning tree $T_C$ on each $C\in\mathcal C$, satisfying the following properties:
    \begin{itemize}
        \item[(i)] The number of clusters in $\mathcal C$ is $O(n/b)+1$.
        \item[(ii)] $\mathbb E\bigl[\sum_{C\in\mathcal C}\omega(T_C)\bigr]\leq \lambda\cdot \opt(P)$.
    \end{itemize}
    The first property holds for every outcome of the random choices.
\end{definition}

A main technical contribution of this section is a randomized MPC construction of a light partition, which is stated in the following lemma.

\begin{lemma}\label{lem:balanced-light-partition-algorithm}
    Fix a constant $\sigma\in (0,1)$.
    Let $n$ and $m$ be two integers with $m\leq n$, and let $b=n^{\sigma/10}$.
    Given a set $P$ of $m$ points in the Euclidean space of dimension $d$, one can construct a $(b,O(1/\sigma^3))$-randomized light partition of $P$ with high probability in $O(1/\sigma)$ MPC rounds, using $O(n^\sigma)$ local space per machine and $O(md+n^{1+\sigma})$ total space.
\end{lemma}
\begin{lemma}\label{lem:balanced-light-partition-algorithm-doubling}
    Fix a constant $\sigma\in (0,1)$.
    Let $n$ and $m$ be two integers with $m\leq n$, and let $b=n^{\sigma/10}$.
    Given a set $P$ of $m$ points in a doubling metric space of doubling dimension $\ddim$ where $\sigma\log n\geq \Omega(\ddim)$, one can construct a $(b,O(1))$-randomized light partition of $P$ with high probability in $O(\ddim/\sigma^3)$ MPC rounds, using $O(n^\sigma)$ local space per machine and $\tilde O(n^{1+\sigma})\cdot 2^{O(\ddim)}$ total space.
\end{lemma}

We first prove Theorem~\ref{thm:euclidean-mst} and Theorem~\ref{thm:doubling-mst} using Lemma~\ref{lem:balanced-light-partition-algorithm} and Lemma~\ref{lem:balanced-light-partition-algorithm-doubling}.
To simplify the presentation, we describe only the Euclidean case.
Notice that Theorem~\ref{thm:euclidean-mst} depends only on a light partition and the triangle inequality.
The doubling metric case follows directly, and thus we defer the proof of Theorem~\ref{thm:doubling-mst} to Appendix~\ref{apd:doubling-proof}.
We then describe the construction of a light partition and its MPC implementation in Section~\ref{sec:construct-BLP}.

\begin{proof}[Proof of Theorem~\ref{thm:euclidean-mst} assuming Lemma~\ref{lem:balanced-light-partition-algorithm}]
We first describe our algorithm.
For pseudocode, see Algorithm~\ref{alg:MST-with-BLP}.
Let $L=\lceil 20/\sigma\rceil$, and keep $\EMPH{b}=n^{\sigma/10}$ fixed throughout the algorithm.
Starting with $P=Q_0$, we repeatedly compute a $(b,O(1/\sigma^3))$-randomized light partition.
At level $i$, we are given a set $Q_i\subseteq P$ of representative points.
Let $(\mathcal C_i, \{T_C\}_{C\in \mathcal C_i})$ be a $(b,O(1/\sigma^3))$-randomized light partition of $Q_i$ obtained by using Lemma~\ref{lem:balanced-light-partition-algorithm}.
We choose an arbitrary representative point per cluster in $\mathcal C_i$, and let $Q_{i+1}$ be the set of chosen representative points, which will be passed to the next level.
Once $n^{\sigma/2}$ representative points remain, we can construct their MST on a single machine.
On every successful call, the number of points becomes $O(m/b)+1$; while $m>n^{\sigma/2}$, this is at most $m/n^{\sigma/20}$ for sufficiently large $n$.
Thus $L$ successful calls suffice.
The union of the trees constructed at all levels forms a spanning tree on $P$. We denote it by $T$.

\begin{algorithm}[t]
    \caption{Metric MST using light partitions}
    \label{alg:MST-with-BLP}
    \small
    \SetKwInput{KwIn}{Input}
    \SetKwInput{KwOut}{Output}
    \KwIn{A set $P$ of $n$ points in a metric space $(X,\dist)$}
    $b\gets n^{\sigma/10}$\;
    $Q_0\gets P$, $E\gets\emptyset$, $i\gets 0$\;
    \While{$|Q_i|>n^{\sigma/2}$}{
        Compute a $(b,\lambda)$-light partition $(\mathcal C_i,\{T_C\}_{C\in\mathcal C_i})$ of $Q_i$\;
        \ForEach{$C\in\mathcal C_i$}{
            $E\gets E\cup E(T_C)$; choose a representative $r_C\in C$\;
        }
        $Q_{i+1}\gets\{r_C:C\in\mathcal C_i\}$\;
        $i\gets i+1$\;
    }
    Compute an MST of $Q_i$ on one machine and add its edges to $E$\;
    \Return{$(P,E)$}\;
\end{algorithm}

We now show that $T$ is an $O(1)$-approximate MST in expectation.
Suppose that there are $\ell<L$ levels.
For each $0\le i\le \ell$, let $m_i=|Q_i|$.
For every $i<\ell$, the trees constructed within the clusters in $\mathcal C_i$ form a spanning forest on $Q_i$, and $Q_{i+1}$ contains exactly one representative from each tree.
Therefore, adding the spanning forest of $Q_{i+1}$ merges those trees in the forest of $Q_i$ without creating a cycle, and thus $T$ is a spanning tree of $P$.
Note that $|E(T)|=\sum^{\ell-1}_{i=0}(\sum_{C\in \mathcal C_i}(|C|-1))+(m_\ell -1)=\sum^{\ell-1}_{i=0}(m_i-m_{i+1})+(m_\ell -1)=n-1$.

Consider an MST $T^*$ of $P$ and an Euler tour of the multigraph obtained by duplicating all edges of $T^*$.
The length of the Euler tour is $2\cdot \omega(T^*)$.
For each $0\le i\le \ell$, let $(q_1,q_2,\ldots ,q_{m_i})$ be the points in $Q_i$ ordered by their first occurrence in the Euler tour.
Then, $\sum_{j=1}^{m_i}\dist(q_j,q_{j+1})\leq 2\cdot \omega(T^*)$ by the triangle inequality, where $q_{m_i+1}=q_1$.
This implies that $\opt(Q_i)\leq 2\cdot \omega(T^*)=2\opt(P)$.
At each executed level, the second property of a randomized light partition bounds the expected total weight of the cluster trees by $O((1/\sigma^3)\opt(Q_i))$, conditional on the preceding iterations and using fresh random choices at that level.
Thus, summing over the $O(1/\sigma)$ possible levels and including the final MST, we have
\[
\begin{aligned}
    \mathbb E[\omega(T)]&=\mathbb E\Bigl[\sum_{i=0}^{\ell-1}\sum_{C\in \mathcal C_i}\omega(T_C)+\opt(Q_\ell)\Bigr]\\
    &\leq O((1/\sigma^4)\cdot \opt(P))+2\opt(P)=O((1/\sigma^4)\cdot \opt(P)).
\end{aligned}
\]

Each iteration constructs a randomized light partition in $O(1/\sigma)$ MPC rounds.
Moreover, we can select representatives locally and update the representative and edge records in $O(1/\sigma)$ MPC rounds.
The MST on $Q_\ell$ is computed locally.
Since there are $\ell=O(1/\sigma)$ iterations, the algorithm takes $O(1/\sigma^2)$ MPC rounds in total.

By the Markov inequality, the probability that $\omega(T)>2c\cdot \opt(P)$ is at most $1/2$ for some constant $c$.
Thus, by applying Algorithm~\ref{alg:MST-with-BLP} independently to $O(\log n)$ copies of $P$ in parallel and returning the tree of minimum total weight, we can construct an $O(1)$-approximate MST with high probability.
By reusing space across iterations and applying the above algorithm with $\sigma/2$ local memory parameter, the bounds in Lemma~\ref{lem:balanced-light-partition-algorithm} also give $n^{\sigma/2}$ local space per machine, and $O(n^{1+\sigma/2})$ total space.
The parallel copies and the selection take $O(1/\sigma)$ MPC rounds and $O(n^{1+\sigma/2}\log n)=O(n^{1+\sigma})$ total space.
\end{proof}

\subsection{Proof of Lemma~\ref{lem:balanced-light-partition-algorithm} and Lemma~\ref{lem:balanced-light-partition-algorithm-doubling}: Computing a light partition}\label{sec:construct-BLP}
In this subsection, we prove Lemma~\ref{lem:balanced-light-partition-algorithm} and Lemma~\ref{lem:balanced-light-partition-algorithm-doubling} by showing how to compute a $(b,\lambda)$-light partition of a point set, with $\lambda=O(1/\sigma^3)$ in $d$-dimensional Euclidean space and $\lambda=O(1)$ in a metric space of doubling dimension $\ddim$.
Our approach is similar to that used to construct a light contraction in the general metric case.
However, unlike in the general metric case, the input consists of the points rather than pairwise distances, making it hard to compute exact balls centered at the points in a constant number of MPC rounds.
We therefore use consistent hashing to approximate the corresponding radii; see Section~\ref{sec:prelim}.
Since these approximate radii need not satisfy the property required by \textsc{ForestCutting}, we define a new value for each point $x\in P$ that satisfies this property.
Moreover, \textsc{Merging-via-Leader-Compression}, which we use in the general metric case, cannot be directly applied here in a constant number of MPC rounds since we cannot compute the nearest root without consistent hashing.
Nevertheless, \textsc{Merging-via-Sampling} suffices to obtain a constant approximation factor.

Throughout this subsection, we are given a set $P$ of $m$ points in a metric space with distance $\dist$.
We first describe how to construct a light partition on $P$ and how to implement this using consistent hashing.
Moreover, we assume \EMPH{polynomial spread}: the ratio of the largest to the smallest nonzero pairwise distance is polynomial in $n$. We show how to get rid of this assumption in Appendix~\ref{apd:spread}.
We also assume that the minimum nonzero distance is at least one; if this is not the case, we can normalize the distances in $O(1/\sigma)$ MPC rounds.
Let $\EMPH{k}=\lceil c b \log n\rceil=\Theta(b\log n)$, where $c$ is a sufficiently large fixed constant.
If $m\leq k$, we compute an exact MST of $P$ on a single machine and return it as a single cluster.

We first construct a light spanning forest using \textsc{TruncatedPrim} on all points in $P$.
In the general metric case, we compute, for each point $x$, a radius such that the ball centered at $x$ with this radius contains at most $k$ points, and apply the exact ball to \textsc{TruncatedPrim}.
However, in this case, it is unclear how to compute the exact ball centered at every point in constant MPC rounds.
Therefore, we use the estimate of this radius and the set of points that approximates the exact ball.
Let $\textcolor{BrickRed}{r_x}$ for every point $x\in P$ be the smallest radius such that $|B(x,r)\cap P|\geq k$, i.e., $r_x=\min\{r:|B(x,r)\cap P|\geq k\}$.
We use estimates $\textcolor{BrickRed}{\hat r_x}$ such that $r_x\leq \hat r_x\leq 2\gamma r_x$, where $\textcolor{BrickRed}{\gamma} >1$ is a constant determined later.

For each point $x\in P$, suppose that we have $\hat r_x$ and a set, denoted by $\textcolor{BrickRed}{S_x}$, of points containing $B(x,\varepsilon \hat r_x)\cap P$ and contained in $B(x,\gamma \varepsilon \hat r_x)\cap P$, where $\textcolor{BrickRed}{\varepsilon}=\frac{1}{4\gamma^2}$ is a constant.
Note that $|S_x|\leq k=O(b\log n)=O(n^{\sigma/2})$ since $2\varepsilon\gamma^2 <1$.
For every point $x\in P$, we suppose that $S_x$ is stored on a single machine.

We then run \textsc{TruncatedPrim}$(x,S_x,\varepsilon\hat r_x/2,0)$ on the machine $M_x$ for every point $x\in P$.
Let $\EMPH{F}$ be the union of the edge sets returned by the calls of \textsc{TruncatedPrim}.
As in the general metric case, we remove the duplicated edges of $F$ in $O(1/\sigma)$ MPC rounds using \textsc{Sorting}.
Since for every $x\in P$, $S_x$ contains $B(x,\varepsilon\hat r_x)$, the resulting edge set $E_x$ is a subset of the MST from Prim's algorithm by Lemma~\ref{lem:truncated-prim}.
Therefore, $F\subseteq T$ and $\omega(F)=O(\opt(P))$.
Moreover, condition (ii) holds immediately by taking a partition as the vertex sets of the connected components of $(P,E(F))$ and each spanning tree in a cluster as the corresponding component tree.

However, we still have two issues: (i) we need to identify the component of $F$, and (ii) there may still be too many components in $F$.
To resolve the first issue, we apply \textsc{ForestCutting} described in Section~\ref{sec:subroutine}, and to resolve the second issue, we merge several components so that we have $O(m/b)+1$ components at the end.

\subsubsection{Forest cutting}
We first apply $\textsc{ForestCutting}$ on $F$.
However, $\hat r_x$ does not necessarily satisfy the assumption of Lemma~\ref{lem:forestcutting}, although the exact radius $r_x$ does.
We therefore define a local radius $\textcolor{BrickRed}{h_x}=\min_{y\in P}\{\dist(x,y)+\frac{\varepsilon}{100}\hat r_y\}$ for each point $x\in P$.
To compute $h_x$ for each point $x\in P$, it suffices to check the points in $S_x$, since $h_x\leq (\varepsilon/100)\cdot \hat r_x$.
Recall that $S_x$ is stored in a single machine, and thus we can compute $h_x$ only using local computation.

We then apply \textsc{ForestCutting} on $F$ with $a_v:=h_v/8$ and $\ell_{uv}:=\dist(u,v)$ for every vertex $v\in V(F)$ and edge $(u,v)\in E(F)$ in $O(1/\sigma)$ MPC rounds.
Recall that the subroutine \textsc{ForestCutting} requires that for a point $x$, $a_x\leq a_y+\dist_F(x,y)/8$, where $y$ is a point in the same tree of $F$.
This holds since the following lemma implies that $h_x/8\leq h_y/8+\dist_F(x,y)/8$.

\begin{lemma}\label{lem:local-radius}
    For any two points $x,y\in P$, $|h_x-h_y|\leq \dist(x,y)$.
    Moreover, for every $x\in P$, $\frac{\varepsilon}{100}r_x\leq h_x\leq \frac{\varepsilon\gamma}{50}r_x$, and $\varepsilon\hat r_x\geq 100h_x$.
\end{lemma}
\begin{proof}
    For a candidate $z\in P$, $|\dist(x,z)+\varepsilon\hat r_z/100-(\dist(y,z)+\varepsilon\hat r_z/100)|=|\dist(x,z)-\dist(y,z)|\leq \dist(x,y)$.
    Let $z^*$ be the point realizing $h_y$.
    We then have $h_x\leq \dist(x,z^*)+\varepsilon\hat r_{z^*}/100\leq h_y+\dist(x,y)$.

    We now show the upper and lower bounds of $h_x$ for each $x\in P$.
    Note that $h_x\leq \dist(x,x)+\varepsilon\hat r_x/100\leq \varepsilon\gamma r_x/50$ since $\hat r_x\leq 2\gamma r_x$, and thus $100 h_x\leq \varepsilon\hat r_x$.
    For any point $y\in P$, we have $r_y\geq r_x-\dist(x,y)$, since the ball of radius $r_y$ centered at $y$ is contained in the ball of radius $r_y+\dist(x,y)$ centered at $x$.
    We then have $\dist (x,y)+\varepsilon\hat r_y/100\geq \dist(x,y)+\varepsilon r_y/100\geq \varepsilon r_x/100+(1-\varepsilon/100)\cdot \dist(x,y)\geq \varepsilon r_x/100$, since $\varepsilon<1$.
    Thus, $h_x\geq \varepsilon r_x/100$.
\end{proof}

Moreover, \textsc{ForestCutting} requires that, for each $x\in V(F)$, the subgraph $F_x$ induced by $\{y\in V(F):\dist_F(x,y)\leq h_x/8\}$ be stored on a single machine.
By Lemma~\ref{lem:local-radius} and our choice of $\varepsilon<1/(2\gamma^2)$, we have $h_x/8<r_x$, so $F_x$ contains at most $k=O(n^\sigma)$ vertices.
Furthermore, since $h_x/8\leq \varepsilon\hat r_x/800$, every vertex of $F_x$ belongs to $S_x$, which is stored on a single machine.
We locally compute an MST $T_x$ of $S_x$ using the same edge ordering as in \textsc{TruncatedPrim}.
Since $F[S_x]\subseteq T_x$, we identify the edges of $F[S_x]$ using \textsc{Aggregation} and \textsc{Broadcasting}, and then compute $F_x$ locally.
These steps take $O(1/\sigma)$ MPC rounds within the stated space bounds.

We now have a low-cost forest partition $\mathcal F$ of $F$ consisting of oriented trees such that
\begin{equation}\label{equ:euclidean-low-cost}
    \mathbb E\left[\sum_{z\in R}h_z/8\right]\leq \sum_{C\in \textsc{cc}(F)}\min_{v\in C}h_v/8+O\left(\sum_{(u,v)\in E(F)}\dist(u,v)\right),
\end{equation}

where $\textsc{cc}(F)$ denotes the connected components of $F$ before any edges are removed and $R$ denotes the roots of the trees of $\mathcal F$, by Definition~\ref{def:low-cost-forest-partition} and Lemma~\ref{lem:forestcutting}.
Moreover, each vertex $v$ has the identifier of its root $\rt(v)$ of the tree containing $v$ in $\mathcal F$, which allows us to identify the component of $\mathcal F$ containing $v$.
Also, condition (ii) still holds, since the edges retained by \textsc{ForestCutting} are contained in an MST of $P$.

The following lemma will be used to show that condition (ii) continues to hold after the second step.
Since its proof follows the same argument as that of Lemma~\ref{lem:general-connecting-cost} in the general metric case, we defer it to Appendix~\ref{apd:h-sum}.

\begin{restatable}{lemma}{euclideanLocalRad}
\label{lem:partitioned-tree-sum-local-rad}
 $\mathbb E[\sum_{z\in R}h_z]=O(\opt(P))$, where $R$ denotes the set of the roots of the trees of $\mathcal F$.
\end{restatable}

\subsubsection{Forest merging}
We now merge components of $\mathcal F$ into $O(m/b)+1$ components by adding edges, while keeping the approximation factor.
For a root $z\in R$, we let $T_z$ be the component in $\mathcal F$ rooted at $z$.

We first sample independently each root $z\in R$ with probability $\min\{1,|T_z|/2^{\lceil \log b\rceil}\}$.
These sampled roots will determine how components in $\mathcal F$ are merged.
Recall that each vertex $v$ has the identifier of its root $\rt(v)$ of the tree containing $v$ in $\mathcal F$, and thus we can compute $|T_z|$ in $O(1/\sigma)$ MPC rounds using \textsc{Aggregation}.
We then sample $z$ independently using local computation.
The expected number of initially sampled components is at most $m/b$, since $\sum_{z\in R}|T_z|=m$.
If no root is sampled, we choose an arbitrary root and treat it as sampled.
For each non-sampled root $z$, suppose that we have a $2\gamma$-approximate nearest sampled root, denoted by \EMPH{$\textsc{ann}(z)$}.
For each sampled root $z$, we set \EMPH{$\textsc{ann}(z)=z$}.

For every sampled root $z$, we add the edges between the roots whose $2\gamma$-approximate nearest sampled root is $z$ as follows.
Let $(T_1,T_2,\ldots,T_\ell)$ be the components in $\mathcal F$ whose roots are assigned to $z$, ordered by lexicographic order.
For every $1\leq i\leq \ell$, we compute the prefix sum $s_i=\sum_{j<i}|T_{j}|$ in $O(1/\sigma)$ rounds using \textsc{PrefixSum}.
We form clusters by grouping together the trees $T_{i}$ with the same value of $\lfloor 2s_i/b^4\rfloor$.
Let $\mathcal C$ be the resulting collection of clusters.

We now construct $T_C$ for each cluster $C\in \mathcal C$.
Let $z(C)$ be the sampled root used to construct $C$.
Note that $C$ may not contain $z(C)$.
We first compute the root $q$ in $C$ closest to $z(C)$.
This can be done in $O(1/\sigma)$ MPC rounds, since the roots in the cluster already have $z(C)$.
We then add the edge between $q$ and all the other roots in $C$ in $O(1/\sigma)$ MPC rounds by sending $q$ to them.
We can do this in parallel for all clusters.

By the following lemma, we can conclude Lemma~\ref{lem:balanced-light-partition-algorithm} in the same way as in the proof of Lemma~\ref{lem:light-contraction}.

\begin{lemma}\label{lem:euclidean-total-added-weight}
    The expected total weight of added edges is $O((\gamma /\varepsilon)\opt(P))$.
\end{lemma}
\begin{proof}
    Let $\mathcal E$ be the event that, for every non-sampled root $z$, there exists at least one sampled root $q$ such that $T_q$ intersects $B(z,r_z)$.
    For each root $z$, the probability that there is no sampled root whose component intersects $B(z,r_z)$ is at most
    \[
        \prod_{x\in R:T_x\cap B(z,r_z)\neq \emptyset}\left(1-p_x\right)\leq \exp(-|B(z,r_z)\cap P|/2^{\lceil \log_2 b\rceil})\leq \exp(-(c/2)\log n).
    \]
    The first inequality holds since the size sum of the components intersecting $B(z,r_z)$ is at least $|B(z,r_z)\cap P|$, and the second inequality holds since $|B(z,r_z)\cap P|\geq k$ and $k=\lceil c b \log n\rceil$.
    A union bound over at most $m\leq n$ roots gives $\Pr[\mathcal E^c]\leq n^{1-c/2}$.

    Fix an outcome in $\mathcal E$ and a non-sampled root $z$.
    There is a sampled root $z'$ such that $T_{z'}$ contains a vertex $v\in B(z,r_z)$.
    Therefore, 
    \[
        \begin{aligned}
            \dist(z,z')&\leq \dist(z,v)+\dist(v,z') \leq r_z + \frac{h_v}{8} \leq r_z+\frac{r_v}{4} \leq \frac{3}{2}r_z\leq 2r_z.
        \end{aligned}
    \]
    The second inequality holds by Lemma~\ref{lem:forestcutting-max-path-length}, and the third inequality holds by Lemma~\ref{lem:local-radius}.
    The last inequality holds since for a point $v$ in $B(z,r_z)$, $r_v\leq r_z+\dist(z,v)\leq 2r_z$.

    In the case that every root $z\in R$ satisfies that $\dist(z,\textsc{ann}(z))\leq 4\gamma r_z$, consider a cluster $C\in \mathcal C$.
    Let $q$ be the root in $C$ closest to $z(C)$.
    By the choice of $q$, every other root $x$ in $C$ satisfies $\dist(x,q)\leq \dist(x,z(C))+\dist(z(C),q)\leq 2\cdot \dist(x,z(C))$.
    Therefore, the total weight of added edges for all clusters is at most $2\cdot \sum_{z\in R}\dist(z,\textsc{ann}(z))\leq 8\gamma  \cdot \sum_{z\in R} r_z\leq (800\gamma /\varepsilon)\sum_{z\in R}h_z$ by Lemma~\ref{lem:local-radius}.
    
    Otherwise, the total added weight is at most $n\cdot \opt(P)$ when $\mathcal E$ does not hold.
    Therefore, the expected total added weight is at most
    \[
        \frac{800\gamma }{\varepsilon}\mathbb E\left[\sum_{z\in R}h_z\right]+n\cdot \Pr[\mathcal E^c]\cdot \opt(P)\leq O((\gamma /\varepsilon)\opt(P)).\qedhere
    \]
\end{proof}

\subsubsection{MPC implementation using consistent hashing}
The above construction uses $O(1/\sigma)$ MPC rounds in addition to those needed to compute $\hat r_x$ and $S_x$ for every point $x\in P$ for \textsc{TruncatedPrim} and \textsc{ForestCutting}, and $\textsc{ann}(z)$ for every root $z$ of $\mathcal F$ for \textsc{Merging-via-Sampling}.
Now, we describe how to compute them using consistent hashing for Euclidean spaces of dimension $d$ and doubling metric spaces of doubling dimension $\ddim$.
Specifically, we will utilize Lemma~\ref{lem:euclidean-consistent-hashing} and Lemma~\ref{lem:doubilng-consistent-hashing} to construct a $(\Gamma,\Lambda)$-hash for a constant $\Gamma$.
We choose $\gamma=2\Gamma+1$, where $\Gamma$ is the constant from Lemma~\ref{lem:euclidean-consistent-hashing} or Lemma~\ref{lem:doubilng-consistent-hashing}.
Then, combining Lemma~\ref{lem:euclidean-total-added-weight}, we conclude Lemma~\ref{lem:balanced-light-partition-algorithm} and Lemma~\ref{lem:balanced-light-partition-algorithm-doubling}.

\paragraph{Computing $\hat r_x$.}
To compute $\hat r_x$ for every point $x\in P$, we first compute a 2-approximation $D$ of the diameter of $P$ in $O(1/\sigma)$ MPC rounds by computing the maximum distance from an arbitrary point. More specifically, we
choose an arbitrary point and send it to all machines using \textsc{Broadcasting}, and then we can compute $D$ using \textsc{Sorting}.
Let $I=\{2^0,2^1,\ldots,2^{\lceil\log D\rceil}\}$.
Each machine constructs $I$.
Since $|I|=O(\log n)$, $I$ fits in a single machine.

In Euclidean spaces, for every scale $t\in I$, we construct a $(\Gamma,\Lambda)$-hash ${\varphi_t}$ of $P$ with diameter bound $2\Gamma t$ using Lemma~\ref{lem:euclidean-consistent-hashing}.
Here, $\Gamma=O(1/\sigma)$ and $\Lambda=O(n^\sigma)$.
In particular, we create one copy of $P$ for each of the $O(\log n)$ scales and construct all hashes in parallel.
For a scale $t$ and each point $x\in P$, let $N_t(x)=\varphi_t^{-1}(\varphi_t(B(x,t)))\cap P$.
For each point $x\in P$, we compute the minimum scale $t_x=\min\{t\in I: |N_t(x)|\geq k\}$.
To do this, for every $t\in I$, we aggregate the number of points in $N_t(x)$.
Then, we set $\hat r_x:=\gamma t_x$.
Since $N_{t_x}(x)\subseteq B(x,(1+2\Gamma)t_x)\cap P\subseteq B(x,\hat r_x)\cap P$, $|B(x,\hat r_x)\cap P|\geq k$.
Therefore, $r_x\leq \hat r_x$.
Moreover, by the choice of $t_x$, $\hat r_x\leq 2\gamma r_x$.

In doubling metric spaces, we compute $\hat r_x$ similarly.
For every scale $t\in I$, we construct a $(\Gamma,\Lambda)$-hash ${\varphi_t}$ of $P$ with diameter bound $\Gamma t$ and its corresponding approximate neighborhood hash pairs in parallel using Lemma~\ref{lem:doubilng-consistent-hashing}.
Here, $\Gamma=O(1)$ and $\Lambda=2^{O(\ddim/\Gamma)}$ in a doubling metric space.
In this case, we cannot obtain $N_t(x)$ immediately from Lemma~\ref{lem:doubilng-consistent-hashing}.
For each point $x\in P$, we count the number of points in $A_P(x,t)$ in $O(1/\sigma)$ MPC rounds using \textsc{Aggregation}.
Then, we compute the minimum scale $t_x=\min\{t\in I: |A_P(x,t)|\geq k\}$, and set $\hat r_x:=\gamma t_x$.
Since $A_P(x,t_x)\subseteq B(x,2\Gamma t_x)\cap P\subseteq  B(x,\hat r_x)\cap P$, we have $r_x\leq \hat r_x$.
Moreover, by the choice of $t_x$, $\hat r_x\leq 2\gamma r_x$.

\paragraph{Computing $S_x$ and collecting it on a single machine.}
Recall that for a point $x\in P$, ${S_x}$ is the set of points containing $B(x,\varepsilon \hat r_x)\cap P$ and contained in $B(x,\gamma \varepsilon \hat r_x)\cap P$.
We construct $S_x$ similarly to how we compute $\hat r_x$.
In particular, we collect $S_x$ into a single machine.
Let $I'=\varepsilon\gamma I=\{\varepsilon\gamma 2^0,\varepsilon\gamma 2^1,\ldots \}$.
Note that for every point $x\in P$, $\varepsilon\hat r_x\in I'$.

We first construct $S_x$.
In Euclidean spaces, for each $t\in I'$, we construct a $(\Gamma,\Lambda)$-hash $\varphi_t$ with diameter bound $2\Gamma t$.
We set $S_x:=N_{\varepsilon\hat r_x}(x)$.
In doubling metric spaces, for each $t\in I'$, we construct a $(\Gamma,\Lambda)$-hash $\varphi_t$ with diameter bound $\Gamma t$ and its corresponding approximate neighborhood hash pairs.
We set $S_x:=A_P(x,\varepsilon\hat r_x)$.
Note that in both cases, $B(x,\varepsilon \hat r_x)\cap P\subseteq S_x\subseteq B(x,\varepsilon\gamma\hat r_x)\cap P\subseteq B(x,2\varepsilon\gamma^2 r_x)$.
Therefore, $|S_x|\le k$, since $2\varepsilon\gamma^2<1$.

We now collect $S_x$ on a single machine in parallel in $O(1/\sigma)$ MPC rounds, using \textsc{Broadcasting}.
In Euclidean spaces, we assign $O(mkd)=O(n^{1+\sigma/2})$ additional space for the collected sets.
We allocate $O(kd)$ space to each point $x$, and let $M_x$ denote its assigned machine, ensuring that the total allocated space on each machine is $O(n^\sigma)$.
For every $t\in I'$ and $y\in P$, we create a record $((t,\varphi_t(y)),y)$.
For each $x\in P$, we create a request $((\varepsilon\hat r_x,q),M_x)$ for every $q\in\varphi_{\varepsilon\hat r_x}(B(x,\varepsilon\hat r_x))$.
Using \textsc{Aggregation}, we compute the number of points in a cluster and associate this count with every candidate request.
We remove all candidate requests with count zero.
This takes $O(1/\sigma)$ MPC rounds.
For each request $((t,q),M_x)$, we send $y$ to $M_x$ for every record $((t,q),y)$ using \textsc{Broadcasting}.\footnote{\textsc{Broadcasting} requires a unique value for each key. Using \textsc{Sorting}, we rank the points sharing each key $(t,q)$ and assign the key $(t,q,j)$ to the $j$th point. For every $q\in\varphi_{\varepsilon\hat r_x}(B(x,\varepsilon\hat r_x))$, we send the number of points with key $(\varepsilon\hat r_x,q)$ to $M_x$. The machine $M_x$ then expands the corresponding request into one request
for each point index. These steps take $O(1/\sigma)$ MPC rounds within the stated total space bound.}
This collects exactly $S_x$ for each $x$.
Since $|S_x|\le k$, the collection respects the allocated space and takes $O(1/\sigma)$ MPC rounds.

Similar to the Euclidean case, in doubling metric spaces, we assign $O(mk)=O(n^{1+\sigma/2})$ additional space for the collected sets, and allocate $O(k)$ space to each point and create a record $(t,\varphi_{t}(y), y)$ for every point $y\in P$.
For each $x\in P$, we create a request $(\varepsilon\hat r_x,q,x,M_x)$ for every approximate neighborhood hash pair $(x,q)$ of $\varphi_{\varepsilon\hat r_x}$.
Then, we can collect $S_x$ as in the Euclidean case.

\paragraph{Computing nearest sampled root.}
To compute $\textsc{ann}(z)$ for every root $z\in R$, we use a consistent hash $\varphi_t$ for every $t\in I$ as we compute $\hat r_x$.
We first compute the minimum scale $t_z$ such that $N_{t_z}(z)$ in the Euclidean case and $A_P(z,t_z)$ in the doubling metric space contain at least one sampled root.
We then choose an arbitrary sampled root in $N_{t_z}(z)$ or $A_P(z,t_z)$ as $\textsc{ann}(z)$.
Note that $\dist(z,\textsc{ann}(z))\leq \gamma t_z\leq 2\gamma\dist(z,X)$, where $X$ is the set of sampled roots.

\section{Approximate TSP in constant rounds}
\label{sec:tsp}
In this section, we present fully scalable MPC algorithms for constructing an approximate TSP in general metric spaces, Euclidean spaces, and doubling metric spaces.
More specifically, we prove the following theorem.

\begin{theorem}\label{thm:tsp}
    Fix constants $\sigma,\varepsilon\in(0,1)$.
    There are randomized MPC algorithms that return an approximate TSP with the following guarantees.
    \begin{itemize}
        \item For a general metric on $n$ points given by all pairwise distances, the algorithm returns a $(2+\varepsilon)$-approximate TSP tour in $O(\log(1/(\sigma\varepsilon))/\sigma^2)$ rounds, using $n^\sigma$ local space and $\widetilde{O}(n^2)$ total space.

        \item For a set of $n$ points in $\mathbb{R}^d$, the algorithm returns an $O(1/\sigma^4)$-approximate TSP tour in $O(1/\sigma^2)$ rounds, using $(dn)^\sigma$ local space and  $O(nd+n^{1+\sigma})$ total space.

        \item For a set of $n$ points in a metric space of doubling dimension $\ddim$ satisfying $\sigma\log n\geq\Omega(\ddim)$, the algorithm returns an $O(1/\sigma)$-approximate TSP tour in $O(\ddim/\sigma^4)$ rounds, using $n^\sigma$ local space and $\widetilde{O}(n^{1+\sigma})\cdot 2^{O(\ddim)}$ total space.
    \end{itemize}
    Each algorithm succeeds with probability at least $1-n^{-\Omega(1)}$.
\end{theorem}

To this end, we construct an Euler tour of an approximate MST as a directed cycle.
Then, if the MST has an approximation factor $\alpha$, we can construct a $2\alpha$-approximate TSP in $O(1/\sigma)$ MPC rounds by computing the first occurrence in the Euler tour of every point using \textsc{Sorting}.
Therefore, it suffices to construct an Euler tour of an approximate MST while applying the algorithms in Section~\ref{sec:general} and Section~\ref{sec:euclidean}.
Recall that our approximate MST algorithms construct a forest consisting of MST edges and merge its components in a bottom-up manner.
Within each level, after applying \textsc{ForestCutting}, we merge each group of trees in a low-cost forest partition by designating one as the leader and attaching the others directly to it.
Across levels, we merge the spanning trees of the clusters in a light contraction or light partition using the tree edges constructed at the next level.
To merge multiple trees, we will use the algorithm in~\cite{jayaram2024massively}.

\begin{lemma}[\cite{jayaram2024massively}]\label{lem:tour-merge}
    Let $\sigma\in (0,1)$ be a fixed constant.
    Suppose that we are given 
    \begin{itemize}
        \item a partition $\mathcal C$ over a set $V$ of $n$ vertices,
        \item for each cluster in $\mathcal C$, a spanning tree of the cluster and its Euler tour,
        \item a spanning tree $T$, and its Euler tour where each vertex corresponds to a cluster in $\mathcal C$, and an edge between two clusters corresponds to a pair of vertices, one from each cluster.
    \end{itemize}
    Then, there is an MPC algorithm that computes an Euler tour of a spanning tree obtained by merging all spanning trees in the clusters of $\mathcal C$ using the edges of $T$.
    It takes $O(1/\sigma)$ MPC rounds, using $O(n^\sigma)$ local space and $O(n)$ total space.
\end{lemma}

\paragraph{Merging the trees in a low-cost forest partition.}
Consider constructing a light contraction or light partition at a given level.
For each tree in the low-cost forest partition $\mathcal F$ obtained by Lemma~\ref{lem:forestcutting}, we first construct an Euler tour $A$ starting at the root and visiting the children of each vertex in lexicographical order.

We first compute the number $s(v)$ of vertices in the subtree of $\mathcal F$ rooted at each vertex $v$.
Let $P_v$ denote the path from the root $\rt(v)$ to $v$ in $\mathcal F$.
By the assumption in Lemma~\ref{lem:forestcutting}, $F_v=\{u\in V(F):\dist_F(v,u)\leq a_v\}$ is stored in a single machine.
Therefore, $P_v$ is stored in a single machine since $P_v\subseteq F_v$ by the definition of a low-cost forest partition.
For each vertex $u\in P_v$, we construct a record $(v,u)$.
The number of records whose second vertex is $v$ equals $s(v)$.
Hence, we can compute all subtree sizes in $O(1/\sigma)$ MPC rounds using \textsc{Sorting} and \textsc{Aggregation}.

We next determine the position of each directed edge in $A$ using $s(\cdot)$.
For every non-root vertex $v$, let $B(v)$ be the total number of vertices in the subtrees rooted at the siblings before visiting $v$ in $A$.
Specifically, if $c_1,\ldots,c_k$ are the children of a vertex in lexicographical order, then
\[
    B(c_i)=\sum_{j<i}s(c_j).
\]
We compute these values by sorting vertices by their parent and lexicographical order and taking exclusive prefix sums of the subtree sizes $s(\cdot)$ within the children of each vertex.
We then broadcast $s(v)$ and $B(v)$ to every machine storing a neighborhood $F_u$ that contains $v$.
These take $O(1/\sigma)$ MPC rounds by using \textsc{Sorting}, \textsc{PrefixSum}, and \textsc{Broadcasting}.
Suppose that $P_v=(\rt(v)=v_0,v_1,v_2,\ldots,v_k=v)$.
Note that, after first visiting $v_i$, the tour $A$ visits the subtrees rooted at the siblings preceding $v_{i+1}$ and then moves to $v_{i+1}$, traversing $1+2B(v_{i+1})$ edges in total.
Therefore, the directed edge from the parent of $v$ to $v$ has position $\sum_{u\in P_v\setminus\{\rt(v)\}}(1+2B(u))$ in $A$.
After first visiting $v$, the tour traverses the $2(s(v)-1)$ directed edges within its subtree before returning to its parent.
Consequently, the reverse edge has position $((\sum_{u\in P_v\setminus\{\rt(v)\}}(1+2B(u)))+ 2s(v)-1)$ in $A$.
Since $P_v$ is stored in a single machine, we can determine the position of each edge using local computation.

We now have an Euler tour of each tree in $\mathcal F$.
When we merge some trees by \textsc{Merging-via-Leader-Compression} and \textsc{Merging-via-Sampling}, we choose the leader tree among the trees.
We also choose the edge to merge the leader tree and each other tree.
These edges form a star whose vertices correspond to the trees and whose center corresponds to the leader tree.
Therefore, we can construct an Euler tour of this star in $O(1/\sigma)$ MPC rounds using \textsc{Sorting}.
Hence, we can apply Lemma~\ref{lem:tour-merge} to construct an Euler tour of the tree obtained by \textsc{Merging-via-Leader-Compression} and \textsc{Merging-via-Sampling}.

\begin{lemma}\label{lem:tour-merge-forestcutting}
    Given a low-cost forest partition $\mathcal F$ constructed by Lemma~4.3, we can compute an Euler tour of each tree of $\mathcal F$ in $O(1/\sigma)$ MPC rounds.
    Moreover, these tours can be updated after each merging iteration of \textsc{Merging-via-Leader-Compression} or \textsc{Merging-via-Sampling} in $O(1/\sigma)$ MPC rounds.
\end{lemma}

\paragraph{Merging the trees across the levels.}
We now merge the Euler tours constructed at different levels in a bottom-up manner.
For each current vertex, we maintain an Euler tour of the tree on the original vertices represented by it.
Initially, each original vertex represents a singleton tree with an empty tour.

Consider a cluster $C$ in the light contraction or light partition at a given level.
The vertices of $C$ represent vertex-disjoint trees whose Euler tours have already been computed: in the general metric case, a vertex corresponds to a contracted tree, and in the Euclidean and doubling metric cases, it is the representative point of a tree.
The spanning tree $T_C$ specifies how these trees are joined, and its Euler tour is available from the construction above.
Moreover, for each edge of $T_C$, the endpoints of the connecting edge in the original vertex set are known.
Therefore, we can apply Lemma~\ref{lem:tour-merge} to construct an Euler tour of the tree obtained by adding these connecting edges to the represented trees.
We do this in parallel for all clusters and associate each resulting tour with the contracted vertex or representative point at the next level.
Each merging step takes $O(1/\sigma)$ MPC rounds.
Since there are $O(1/\sigma)$ levels, this merging procedure takes $O(1/\sigma^2)$ additional MPC rounds, using $O(n^\sigma)$ local space and $O(n)$ total space.
Therefore, we conclude Theorem~\ref{thm:tsp}.

\section{Conclusion}
\label{sec:conclusion}
In this paper, we present randomized MPC algorithms that construct an approximate MST, with high probability, in general metric spaces, Euclidean spaces of arbitrary dimension, and doubling metric spaces of bounded doubling dimension.
These algorithms improve the round complexities of the best-known MPC algorithms for general metrics and high-dimensional Euclidean spaces~\cite{azarmehr2025massively, jayaram2024massively}.
The round complexity for general metrics matches the dependence on $\varepsilon$ in the known lower bound~\cite{azarmehr2025massively} under the 1-vs-2-cycles conjecture.
Our results also lead to several open questions.

\begin{itemize}
    \item \textbf{$(1+\varepsilon)$-approximate Euclidean MST.} Our algorithm constructs a constant-factor approximate Euclidean MST in $O(1)$ rounds, whereas the MST cost can already be estimated within a factor of $(1+\varepsilon)$ in $O(1)$ rounds using near-linear total space~\cite{10.1145/3564246.3585168}.
    Applying our general-metric algorithm after computing all pairwise distances also yields a $(1+\varepsilon)$-approximate tree, but requires $\widetilde O(nd+n^2)$ total space.
    Can a $(1+\varepsilon)$-approximate Euclidean MST be constructed in $O(1)$ rounds using $(nd)^\sigma$ local space per machine and subquadratic additional total space?
    
    \item \textbf{Near-linear total space for Euclidean spaces.} Even for a constant approximation factor, our constant-round algorithm uses $O(nd+n^{1+\sigma})$ total space.
    Can this be reduced to $\widetilde O(nd)$ while retaining $O(1)$ rounds and $(nd)^\sigma$ local space?
    Our implementation stores overlapping neighborhoods separately to support local forest construction and cutting. 
    
    \item \textbf{Improved approximation for metric Steiner tree.}
    By constructing an approximate MST on the terminals, our algorithms yield fully scalable MPC algorithms for computing a $(2+\varepsilon)$-approximate Steiner tree in general metric spaces and an $O(1)$-approximate Steiner tree in Euclidean and doubling metric spaces~\cite{gilbert1968steiner, mahabadi2025}.
    A recent fully scalable MPC algorithm~\cite{wang2025massively} computes a $(2(1-1/|R|))$-approximate Steiner tree in weighted graphs in $O(\sigma^{-1}\log n+D)$ MPC rounds, using $O(n^{3-\sigma/2})$ total space, where $R$ is the terminal set and $D$ is the shortest-path diameter.
    However, in the sequential setting, a $(\ln 4+\varepsilon)$-approximation is known~\cite{byrka2013steiner}.
    This leaves open whether a better approximation can be achieved while retaining constant round complexity.
    
    \item \textbf{Exact Euclidean MST with low dimension.} Chang and Zheng~\cite{chang2024planargraph} present a constant-round fully scalable MPC algorithm for constructing an exact Euclidean MST in two dimensions using linear total space.
    As noted in~\cite{chang2024planargraph}, it remains open whether the same guarantees can be achieved even in three dimensions.
    
    \item \textbf{Distributed model.}
    Our algorithms gather overlapping neighborhoods and redistribute component information, which may require sending the same point to several machines.
    Thus, their small local-memory bounds do not by themselves ensure near-linear total messages.
    In the related \textsc{Congested Clique} model, exact MSTs can be computed in constant rounds using $O(n^2)$ messages, when each node initially knows its incident edges and their weights~\cite{nowicki2021deterministic}.
    Pemmaraju and Sardeshmukh~\cite{pemmaraju16} give an exact MST algorithm using $O(\log^* n/\rho)$ rounds and $\widetilde O(n^{1+\rho}/\rho)$ messages, for any $0<\rho\le 1$.
    Moreover, \cite{hegeman2014near} presented a constant-factor approximate MST algorithm for doubling metric spaces with $\ddim=O(1)$, using constant rounds and $O(n^2)$ messages.
    Coordinate input has also been studied: for $n$ distinct points in $\{0,1\}^n$ given one per node, Lingas~\cite{lingas2024boolean} constructs a constant-factor approximate Hamming MST in $O(\log^3 n)$ rounds.     
    Can we construct a constant-factor approximate MST in high-dimensional Euclidean space in the \textsc{Congested Clique} model using constant rounds and $\widetilde O(n)$ messages when each node initially holds only the coordinates of one point?

\end{itemize}

\section*{AI methodology}
The authors developed the ideas underlying Theorem~\ref{thm:general-metric-mst} and Theorem~\ref{thm:euclidean-mst} through discussions with ChatGPT-6 Astra. They also used ChatGPT-6 Astra to improve the writing of the paper. The authors verified the correctness and originality of all content, including references.

\bibliographystyle{alphaurl}
\bibliography{papers}

@inproceedings{andoni2014parallel,
  title={Parallel algorithms for geometric graph problems},
  author={Andoni, Alexandr and Nikolov, Aleksandar and Onak, Krzysztof and Yaroslavtsev, Grigory},
  booktitle={Proceedings of the 46th Annual ACM Symposium on Theory of Computing (STOC 2014)},
  pages={574--583},
  year={2014},
  doi = {10.1145/2591796.2591805}
}

@inproceedings{yaroslavtsev2018massively,
  title={Massively parallel algorithms and hardness for single-linkage clustering under {$\ell_p$} distances},
  author={Yaroslavtsev, Grigory and Vadapalli, Adithya},
  booktitle={Proceedings of the 35th International Conference on Machine Learning (ICML 2018)},
  pages={5600--5609},
  year={2018},
  url = {https://proceedings.mlr.press/v80/yaroslavtsev18a.html}
}

@inproceedings{mordacq2026t,
  title={{T-REGS}: minimum spanning tree regularization for self-supervised learning},
  author={Mordacq, Julie and Loiseaux, David and Kalogeiton, Vicky and Oudot, Steve},
  booktitle={Proceedings of the 39th Annual Conference on Neural Information Processing Systems (NeurIPS 2025)},
  pages={12184--12214},
  year={2025},
  doi={10.52202/085713-0408}
}

@inproceedings{march2010fast,
  title={Fast {Euclidean} minimum spanning tree: algorithm, analysis, and applications},
  author={March, William B. and Ram, Parikshit and Gray, Alexander G.},
  booktitle={Proceedings of the 16th ACM SIGKDD International Conference on Knowledge Discovery and Data Mining (KDD 2010)},
  pages={603--612},
  year={2010},
  doi={10.1145/1835804.1835882}
}

@article{yang2020glamst,
  title={{GLaMST}: grow lineages along minimum spanning tree for B cell receptor sequencing data},
  author={Yang, Xingyu and Tipton, Christopher M. and Woodruff, Matthew C. and Zhou, Enlu and Lee, F. Eun-Hyung and Sanz, I{\~n}aki and Qiu, Peng},
  journal={BMC Genomics},
  volume={21},
  number={Suppl 9},
  pages={583},
  year={2020},
  doi={10.1186/s12864-020-06936-w}
}

@InProceedings{driemel_et_al:LIPIcs.SoCG.2025.42,
  author =	{Driemel, Anne and Monemizadeh, Morteza and Oh, Eunjin and Staals, Frank and Woodruff, David P.},
  title =	{Range Counting Oracles for Geometric Problems},
  booktitle =	{Proceedings of the 41st International Symposium on Computational Geometry (SoCG 2025)},
  pages =	{42:1--42:16},
  year =	{2025},
  doi =		{10.4230/LIPIcs.SoCG.2025.42},
}

@inproceedings{DBLP:conf/soda/BalliuLMOU23,
  author       = {Alkida Balliu and
                  Rustam Latypov and
                  Yannic Maus and
                  Dennis Olivetti and
                  Jara Uitto},
  title        = {Optimal deterministic massively parallel connectivity on forests},
  booktitle    = {Proceedings of the 2023 Annual {ACM-SIAM} Symposium on Discrete Algorithms (SODA 2023)},
  pages        = {2589--2631},
  year         = {2023},
  doi          = {10.1137/1.9781611977554.CH99},
}

@article{esfandiari2020parallel,
author = {Behnezhad, Soheil and Dhulipala, Laxman and Esfandiari, Hossein and {\L}{\k{a}}cki, Jakub and Mirrokni, Vahab and Schudy, Warren},
title = {Parallel graph algorithms in constant adaptive rounds: theory meets practice},
year = {2020},
volume = {13},
number = {13},
issn = {2150-8097},
doi = {10.14778/3424573.3424579},
journal = {Proceedings of the VLDB Endowment},
pages = {3588–3602},
numpages = {15}
}

@article{czumaj2005approximating,
  title={Approximating the weight of the {E}uclidean minimum spanning tree in sublinear time},
  author={Czumaj, Artur and Erg{\"u}n, Funda and Fortnow, Lance and Magen, Avner and Newman, Ilan and Rubinfeld, Ronitt and Sohler, Christian},
  journal={SIAM Journal on Computing},
  volume={35},
  number={1},
  pages={91--109},
  year={2005},
  doi={10.1137/S0097539703435297}
}

@inproceedings{patlin2025sublinear,
  title={Sublinear-time algorithm for {MST}-weight revisited},
  author={Patlin, Gryphon and van den Brand, Jan},
  booktitle={Proceedings of the 2025 Symposium on Simplicity in Algorithms (SOSA 2025)},
  pages={46--53},
  year={2025},
  doi={10.1137/1.9781611978315.3}
}

@article{czumaj2009estimating,
  title={Estimating the weight of metric minimum spanning trees in sublinear time},
  author={Czumaj, Artur and Sohler, Christian},
  journal={SIAM Journal on Computing},
  volume={39},
  number={3},
  pages={904--922},
  year={2009},
  doi={10.1137/060672121}
}

@article{chazelle2005approximating,
  title={Approximating the minimum spanning tree weight in sublinear time},
  author={Chazelle, Bernard and Rubinfeld, Ronitt and Trevisan, Luca},
  journal={SIAM Journal on Computing},
  volume={34},
  number={6},
  pages={1370--1379},
  year={2005},
  doi={10.1137/S0097539702403244}
}

@misc{jiang2026round,
  title={Round-efficient fully-scalable {MPC} algorithms for k-Means},
  author={Jiang, Shaofeng H.-C. and Jin, Yaonan and Lou, Jianing and Wang, Weicheng},
  year={2026},
  eprint={2604.00954},
  archivePrefix={arXiv},
  doi={10.48550/arXiv.2604.00954}
}

@article{lai2009approximate,
  title={Approximate minimum spanning tree clustering in high-dimensional space},
  author={Lai, Chih and Rafa, Taras and Nelson, Dwight E.},
  journal={Intelligent Data Analysis},
  volume={13},
  number={4},
  pages={575--597},
  year={2009},
  doi={10.3233/IDA-2009-0382}
}

@InProceedings{pmlr-v119-moor20a,
  title = 	 {Topological autoencoders},
  author =       {Moor, Michael and Horn, Max and Rieck, Bastian and Borgwardt, Karsten},
  booktitle = 	 {Proceedings of the 37th International Conference on Machine Learning (ICML 2020)},
  pages = 	 {7045--7054},
  year = 	 {2020},
  url = 	 {https://proceedings.mlr.press/v119/moor20a.html}
}

@InProceedings{pmlr-v119-hofer20a,
  title = 	 {Topologically densified distributions},
  author =       {Hofer, Christoph and Graf, Florian and Niethammer, Marc and Kwitt, Roland},
  booktitle = 	 {Proceedings of the 37th International Conference on Machine Learning (ICML 2020)},
  pages = 	 {4304--4313},
  year = 	 {2020},
  url = 	 {https://proceedings.mlr.press/v119/hofer20a.html}
}

@article{10.1093/mnras/stt2136,
    author = {Alpaslan, Mehmet and Robotham, Aaron S. G. and Driver, Simon and Norberg, Peder and Baldry, Ivan and Bauer, Amanda E. and Bland-Hawthorn, Joss and Brown, Michael and Cluver, Michelle and Colless, Matthew and Foster, Caroline and Hopkins, Andrew and Van Kampen, Eelco and Kelvin, Lee and Lara-Lopez, Maritza A. and Liske, Jochen and Lopez-Sanchez, Angel R. and Loveday, Jon and McNaught-Roberts, Tamsyn and Merson, Alexander and Pimbblet, Kevin},
    title = {{Galaxy And Mass Assembly (GAMA)}: the large-scale structure of galaxies and comparison to mock universes},
    journal = {Monthly Notices of the Royal Astronomical Society},
    volume = {438},
    number = {1},
    pages = {177-194},
    year = {2014},
    doi = {10.1093/mnras/stt2136}
}

@article{trapnell2014pseudo,
  title={The dynamics and regulators of cell fate decisions are revealed by pseudotemporal ordering of single cells},
  author={Trapnell, Cole and Cacchiarelli, Davide and Grimsby, Jonna and Pokharel, Prapti and Li, Shuqiang and Morse, Michael and Lennon, Niall J. and Livak, Kenneth J. and Mikkelsen, Tarjei S. and Rinn, John L.},
  journal={Nature Biotechnology},
  volume={32},
  number={4},
  year={2014},
  pages = {381--386},
    doi   = {10.1038/nbt.2859}
}

@article{sammut2024predictability,
  title={Predictability of {B} cell clonal persistence and immunosurveillance in breast cancer},
  author  = {Sammut, Stephen-John and
             Galson, Jacob D. and
             Minter, Ralph and
             Sun, Bo and
             Chin, Suet-Feung and
             De Mattos-Arruda, Leticia and
             Finch, Donna K. and
             Sch{\"a}tzle, Sebastian and
             Dias, Jorge and
             Rueda, Oscar M. and
             Seoane, Joan and
             Osbourn, Jane and
             Caldas, Carlos and
             Bashford-Rogers, Rachael J. M.},
  journal={Nature Immunology},
  volume={25},
  number={5},
  pages={916--924},
  year={2024},
  doi={10.1038/s41590-024-01821-0}
}

@article{dean2008mapreduce,
  title={{MapReduce}: simplified data processing on large clusters},
  author={Dean, Jeffrey and Ghemawat, Sanjay},
  journal={Communications of the ACM},
  volume={51},
  number={1},
  pages={107--113},
  year={2008},
  doi={10.1145/1327452.1327492}
}

@inproceedings{dean2004mapreduce,
  author    = {Dean, Jeffrey and Ghemawat, Sanjay},
  title     = {{MapReduce}: simplified data processing on large clusters},
  booktitle = {Proceedings of the 6th Symposium on Operating Systems Design and Implementation (OSDI 2004)},
  pages     = {137--150},
  year      = {2004},
  url       = {https://www.usenix.org/conference/osdi-04/mapreduce-simplified-data-processing-large-clusters}
}

@book{white2012hadoop,
  author    = {White, Tom},
  title     = {{Hadoop}: the definitive guide},
  edition   = {3rd},
  year      = {2012},
  publisher={O'Reilly Media, Inc.},
  url       = {https://www.oreilly.com/library/view/hadoop-the-definitive/9781449328917/}
}

@inproceedings{isard2007dryad,
  title={Dryad: distributed data-parallel programs from sequential building blocks},
  author={Isard, Michael and Budiu, Mihai and Yu, Yuan and Birrell, Andrew and Fetterly, Dennis},
  booktitle={Proceedings of the 2nd ACM SIGOPS/EuroSys European Conference on Computer Systems 2007 (EuroSys 2007)},
  pages={59--72},
  year={2007},
  doi={10.1145/1272996.1273005}
}

@inproceedings{karloff2010model,
  title={A model of computation for {MapReduce}},
  author={Karloff, Howard and Suri, Siddharth and Vassilvitskii, Sergei},
  booktitle={Proceedings of the 21st Annual ACM-SIAM Symposium on Discrete Algorithms (SODA 2010)},
  pages={938--948},
  year={2010},
  doi={10.1137/1.9781611973075.76}
}

@inproceedings{assadi2019massively,
  title={Massively parallel algorithms for finding well-connected components in sparse graphs},
  author={Assadi, Sepehr and Sun, Xiaorui and Weinstein, Omri},
  booktitle={Proceedings of the 2019 ACM Symposium on Principles of Distributed Computing (PODC 2019)},
  pages={461--470},
  year={2019},
  doi={10.1145/3293611.3331596}
}

@article{ghaffari2019massively,
  title={Massively parallel algorithms},
  author={Ghaffari, Mohsen},
  journal={Lecture Notes from ETH Zurich},
  url={http://people.csail.mit.edu/ghaffari/MPA19/Notes/MPA.pdf},
  year={2019}
}

@article{coy2022deterministic,
author = {Coy, Sam and Czumaj, Artur},
title = {Deterministic massively parallel connectivity},
journal = {SIAM Journal on Computing},
volume = {52},
number = {5},
pages = {1269-1318},
year = {2023},
doi = {10.1137/22M1520177},
}

@inproceedings{andoni2018parallel,
  title={Parallel graph connectivity in log diameter rounds},
  author={Andoni, Alexandr and Song, Zhao and Stein, Clifford and Wang, Zhengyu and Zhong, Peilin},
  booktitle={Proceedings of the 59th Annual Symposium on Foundations of Computer Science (FOCS 2018)},
  pages={674--685},
  year={2018},
  doi={10.1109/FOCS.2018.00070}
}

@inproceedings{nowicki2021deterministic,
  author    = {Nowicki, Krzysztof},
  title     = {A deterministic algorithm for the {MST} problem in constant rounds of {Congested Clique}},
  booktitle = {Proceedings of the 53rd Annual ACM Symposium on Theory of Computing (STOC 2021)},
  pages     = {1154--1165},
  year      = {2021},
  doi       = {10.1145/3406325.3451136}
}

@inproceedings{chen2021EMD,
author = {Chen, Xi and Jayaram, Rajesh and Levi, Amit and Waingarten, Erik},
title = {New streaming algorithms for high dimensional {EMD and MST}},
year = {2022},
doi = {10.1145/3519935.3519979},
booktitle = {Proceedings of the 54th Annual ACM Symposium on Theory of Computing (STOC 2022)},
pages = {222--233}
}

@InProceedings{gan2025EMST,
  author =	{Gan, Junhao and Wirth, Anthony and Zhang, Zhuo},
  title =	{{$O(1)$-round MPC algorithms for multi-dimensional grid graph connectivity, Euclidean MST and DBSCAN}},
  booktitle =	{Proceedings of the 28th International Conference on Database Theory (ICDT 2025)},
  pages =	{7:1--7:20},
  year =	{2025},
  doi =		{10.4230/LIPIcs.ICDT.2025.7}
}

@inproceedings{chang2024planargraph,
    author = {Yi-Jun Chang and Da Wei Zheng},
    title = {Fully Scalable Massively Parallel Algorithms for Embedded Planar Graphs},
    booktitle = {Proceedings of the 2024 Annual ACM-SIAM Symposium on Discrete Algorithms (SODA 2024)},
    pages = {4410-4450},
    doi = {10.1137/1.9781611977912.155},
    year={2024}
}

@inproceedings{behnezhad2019near,
  title={Near-optimal massively parallel graph connectivity},
  author={Behnezhad, Soheil and Dhulipala, Laxman and Esfandiari, Hossein and {\L}{\k{a}}cki, Jakub and Mirrokni, Vahab},
  booktitle={Proceedings of the 60th Annual Symposium on Foundations of Computer Science (FOCS 2019)},
  pages={1615--1636},
  year={2019},
  doi={10.1109/FOCS.2019.00095}
}

@article{nanongkai2022equivalence,
  title={Equivalence classes and conditional hardness in massively parallel computations},
  author={Nanongkai, Danupon and Scquizzato, Michele},
  journal={Distributed Computing},
  volume={35},
  number={2},
  pages={165--183},
  year={2022},
  doi={10.1007/s00446-021-00418-2 }
}

@inproceedings{10.1145/3564246.3585168,
    author = {Chen, Xi and Cohen-Addad, Vincent and Jayaram, Rajesh and Levi, Amit and Waingarten, Erik},
    title = {Streaming {Euclidean MST} to a constant factor},
    year = {2023},
    doi = {10.1145/3564246.3585168},
    booktitle = {Proceedings of the 55th Annual ACM Symposium on Theory of Computing (STOC 2023)},
    pages = {156--169}
}

@InProceedings{pemmaraju16,
  author =	{Pemmaraju, Sriram V. and Sardeshmukh, Vivek B.},
  title =	{{Super-fast MST algorithms in the {Congested Clique} using $o(m)$ messages}},
  booktitle =	{Proceedings of the 36th IARCS Annual Conference on Foundations of Software Technology and Theoretical Computer Science (FSTTCS 2016)},
  pages =	{47:1--47:15},
  year =	{2016},
  doi =		{10.4230/LIPIcs.FSTTCS.2016.47}
}

@article{gilbert1968steiner,
  title={Steiner minimal trees},
  author={Gilbert, Edgar N. and Pollak, Henry O.},
  journal={SIAM Journal on Applied Mathematics},
  volume={16},
  number={1},
  pages={1--29},
  year={1968},
  doi={10.1137/0116001}
}

@inproceedings{mahabadi2025,
  author =	{Mahabadi, Sepideh and Roghani, Mohammad and Tarnawski, Jakub and Vakilian, Ali},
  title =	{{Sublinear metric Steiner tree via improved bounds for set cover}},
  booktitle =	{Proceedings of the 16th Innovations in Theoretical Computer Science Conference (ITCS 2025)},
  pages =	{74:1--74:24},
  year =	{2025},
  doi =		{10.4230/LIPIcs.ITCS.2025.74}
}

@article{byrka2013steiner,
  title={Steiner tree approximation via iterative randomized rounding},
  author={Byrka, Jaros{\l}aw and Grandoni, Fabrizio and Rothvo{\ss}, Thomas and Sanit{\`a}, Laura},
  journal={Journal of the ACM},
  volume={60},
  number={1},
  pages={6:1--6:33},
  articleno = {6},
  year={2013},
  doi={10.1145/2432622.2432628}
}

@inproceedings{lingas2024boolean,
  title={Boolean matrix multiplication for highly clustered data on the {Congested Clique}},
  author={Lingas, Andrzej},
  booktitle={Proceedings of the 30th European Conference on Parallel Processing (Euro-Par 2024)},
  pages={45--58},
  year={2024},
  doi={10.1007/978-3-031-69583-4_4}
}

@inproceedings{wang2025massively,
  title={Massively parallel approximate {Steiner} tree algorithms},
  author={Wang, Chilei and Hua, Qiang-Sheng and Jin, Hai},
  booktitle={Proceedings of the 31st International Computing and Combinatorics Conference (COCOON 2025)},
  pages={110--124},
  year={2025},
  doi={10.1007/978-981-95-0218-9_9}
}

@inproceedings{hegeman2014near,
  title={Near-constant-time distributed algorithms on a {Congested Clique}},
  author={Hegeman, James W. and Pemmaraju, Sriram V. and Sardeshmukh, Vivek B.},
  booktitle={Proceedings of the 28th International Symposium on Distributed Computing (DISC 2014)},
  pages={514--530},
  year={2014},
  doi={10.1007/978-3-662-45174-8_35}
}

@inproceedings{jayaram2024massively,
  title={Massively parallel algorithms for high-dimensional {Euclidean} minimum spanning tree},
  author={Jayaram, Rajesh and Mirrokni, Vahab and Narayanan, Shyam and Zhong, Peilin},
  booktitle={Proceedings of the 2024 Annual ACM-SIAM Symposium on Discrete Algorithms (SODA 2024)},
  pages={3960--3996},
  year={2024},
  doi={10.1137/1.9781611977912.139}
}

@inproceedings{JohnsonLindenstrauss1984Extensions,
  author    = {Johnson, William B. and Lindenstrauss, Joram},
  title     = {Extensions of {Lipschitz} Mappings into a {Hilbert} Space},
  booktitle = {Conference in Modern Analysis and Probability},
  doi       = {10.1090/conm/026/737400},
  series    = {Contemporary Mathematics},
  volume    = {26},
  pages     = {189--206},
  year      = {1984}
}

@article{Goodrich1999Sorting,
  author  = {Goodrich, Michael T.},
  title   = {Communication-Efficient Parallel Sorting},
  journal = {SIAM Journal on Computing},
  volume  = {29},
  number  = {2},
  pages   = {416--432},
  year    = {1999},
  doi     = {10.1137/S0097539795294141}
}

@inproceedings{GoodrichSitchinavaZhang2011Sorting,
  author    = {Goodrich, Michael T. and Sitchinava, Nodari and Zhang, Qin},
  title     = {Sorting, Searching, and Simulation in the {MapReduce} Framework},
  booktitle = {Proceedings of the 22nd International Symposium on Algorithms and Computation (ISAAC 2011)},
  pages     = {374--383},
  year      = {2011},
  doi       = {10.1007/978-3-642-25591-5_39}
}

@inproceedings{czumaj2022streaming,
  title={Streaming facility location in high dimension via geometric hashing},
  author={Czumaj, Artur and Jiang, Shaofeng H.-C. and Krauthgamer, Robert and Vesel\'{y}, Pavel and Yang, Mingwei},
  booktitle={Proceedings of the 63rd Annual Symposium on Foundations of Computer Science (FOCS 2022)},
  pages={450--461},
  year={2022},
doi={10.1109/FOCS54457.2022.00050}
}

@inproceedings{jia2005universal,
  title={Universal approximations for {TSP}, {S}teiner tree, and set cover},
  author={Jia, Lujun and Lin, Guolong and Noubir, Guevara and Rajaraman, Rajmohan and Sundaram, Ravi},
  booktitle={Proceedings of the 37th Annual ACM Symposium on Theory of Computing (STOC 2005)},
  pages={386--395},
  year={2005},
  doi={10.1145/1060590.1060649}
}

@inproceedings{azarmehr2025massively,
  title={Massively parallel minimum spanning tree in general metric spaces},
  author={Azarmehr, Amir and Behnezhad, Soheil and Jayaram, Rajesh and {\L}{\k{a}}cki, Jakub and Mirrokni, Vahab and Zhong, Peilin},
  booktitle={Proceedings of the 2025 Annual ACM-SIAM Symposium on Discrete Algorithms (SODA 2025)},
  pages={143--174},
  year={2025},
  doi={10.1137/1.9781611978322.5}
}

@misc{CJFKVY23,
      title        = {Streaming facility location in high dimension via geometric hashing},
      author       = {Artur Czumaj and
	Arnold Filtser and
	Shaofeng H.{-}C. Jiang and
	Robert Krauthgamer and
	Pavel Vesel{\'{y}} and
	Mingwei Yang},
      year={2023},
      eprint={2204.02095},
      archivePrefix={arXiv},
      doi={10.48550/arXiv.2204.02095},
      note	     = {Full version; a preliminary version appeared in FOCS 2022},
}

@InProceedings{CGJKV24,
  author =	{Czumaj, Artur and Gao, Guichen and Jiang, Shaofeng H.-C. and Krauthgamer, Robert and Vesel\'{y}, Pavel},
  title =	{{Fully-scalable MPC algorithms for clustering in high dimension}},
  booktitle =	{Proceedings of the 51st International Colloquium on Automata, Languages, and Programming (ICALP 2024)},
  pages =	{50:1--50:20},
  year =	{2024},
  doi =		{10.4230/LIPIcs.ICALP.2024.50},
}

@misc{bhore2026geometrichashingmpcapplications,
      title={Geometric hashing in {MPC} with applications to clustering and stochastic embeddings}, 
      author={Sujoy Bhore and Arnold Filtser and Eunjin Oh and Hyeonjun Shin and Himanshi Singh},
      year={2026},
      eprint={2610.21371},
      archivePrefix={arXiv},
      doi = {10.48550/arXiv.2610.21371},
      note={To appear in SODA 2027}
}

@InProceedings{oh_et_al:LIPIcs.ESA.2026.88,
  author =	{Oh, Eunjin and Shin, Hyeonjun},
  title =	{Fully Scalable {MPC} Algorithms for {WSPD} in Doubling and {E}uclidean Spaces},
  booktitle =	{Proceedings of the 34th Annual European Symposium on Algorithms (ESA 2026)},
  pages =	{88:1--88:16},
  year =	{2026},
  doi =		{10.4230/LIPIcs.ESA.2026.88},
}
\newpage 
\appendix
\section{Getting rid of the assumption on bounded spread}
\label{apd:spread}
In the Euclidean and doubling cases, we assume \EMPH{polynomial spread} on an input point set $P$ of size $n$: the ratio $\Phi$ of the largest to the smallest nonzero pairwise distance is polynomial in $n$. 
The only place where this assumption is used is to
apply consistent hashing.
In particular,
we construct consistent hashes with $O(\log \Phi)$ different radius scales in parallel. Then the space complexity becomes $O(n\log \Phi)$.
In this section, we preprocess $P$ so that this assumption is satisfied. 

\subsection{Euclidean space}
In a $d$-dimensional Euclidean space, we can use a grid as follows.
Fix an arbitrary point $x\in P$, and compute $\Delta:=\max_{p\in P}\dist(p,x)$ in $O(1/\sigma)$ MPC rounds using $O(dn)$ total space. 
Then consider a grid whose cells have diameter $\Delta/(n^2\sqrt d)$.
For every point $p\in P$, we move it to the center of the grid cell containing $p$.
The error induced by the perturbation is at most $\Delta/n\leq \opt(P)/n \leq \eps\cdot \opt(P)$ for sufficiently large $n$.
In this way, the spread of the resulting point set becomes $O(dn^2)$ since its diameter is $O(\Delta)$.
Since $d=O(\log n)$, the resulting point set has polynomial spread.

We retain only one copy of the center of each occupied grid cell.
For each such cell $C$, we also choose an original point $p_C\in P\cap C$ and associate its identifier with every point in $C$.
We apply the algorithm in Section~\ref{sec:euclidean} to the distinct cell centers.
Note that the additional space needed to duplicate the point set across all scales is $O(nd\log(dn^2))\leq O(n^{1+\sigma})$, since $d=O(\log n)$.
Therefore, without increasing the space complexity, we can get rid of the polynomial spread assumption.

To obtain a spanning tree of $P$, we replace each tree edge between two cell centers by the edge between their chosen points, and connect every other point in each cell $C$ to $p_C$.
Choosing these points and constructing the corresponding edges take $O(1/\sigma)$ MPC rounds using \textsc{Sorting}, \textsc{Aggregation}, and \textsc{Broadcasting}, with $O(nd)$ total space.

\subsection{Doubling space}
We show how to preprocess an $n$-point metric $(X,\dist)$ with doubling dimension $d\leq c\sigma\log n$, for a sufficiently small constant $c>0$, so that the resulting point set has spread $O(n^8)$ with high probability. The error induced by the preprocessing is at most $\eps\cdot\opt(X)$ for any fixed approximation factor and sufficiently large $n$. The preprocessing uses $\widetilde O(n)$ total space, $n^\sigma$ local space per machine, and $O(1)$ rounds for fixed $\sigma$. We assume that each point is represented by $O(1)$ words and that its distance to another point can be computed from their representations.

As in the Euclidean case, we assign points to representatives, apply the algorithm to the representatives, and then recover a spanning tree of the original point set.
We fix an arbitrary point $x\in X$, and compute $\Delta:=\max_{p\in X}\dist(x,p)$. Then $\Delta\leq\opt(X)$ and the diameter of $X$ is at most $2\Delta$. Let $\tau=\Delta/n^2$ and $s=n^\sigma$. Recall that $s$ is the size of the local space.
Instead of computing a net, we adapt the construction of a \EMPH{partition cover tree} introduced by Oh and Shin~\cite{oh_et_al:LIPIcs.ESA.2026.88}.

\paragraph{Algorithm.}
Starting with a single piece $X$, we process the remaining pieces in parallel in phases.
For each piece $U$ with $|U|>s$, take $q=\Theta(\beta\sqrt{s}\log^2 n)$ independent uniform samples from $U$, keeping multiplicities, where $\beta=2^{O(d)}$ is a sufficiently large integer. 
Since $q\leq s^{3/4}$, we can store the samples in a single machine.
On this machine, we recursively partition the \emph{sample} using the balanced-ball rule: for the current nonempty sample multiset $Y$, find a smallest-radius ball $B(z,r)$ centered at a point $z$ in $Y$ containing at least $\lceil |Y|/\beta\rceil$ sampled points.
If $r>\tau$, choose the cut radius uniformly from $[r,2r]$ and recurse on both sides.
Otherwise, cut with the ball $B(z,\tau)$, make the inside a final part, and recurse only on the outside.
We do this until every part either becomes a final part, or contains a constant number of sample points.
Finally, we obtain a partition of the sampled points into subsets.
We represent the resulting partition by a recursion tree $T$ whose internal nodes record the centers and radii of the cuts.
Each leaf corresponds to the region determined by the cuts along its root-to-leaf path, including regions containing no sampled points.

We broadcast $T$ to all the machines holding $U$.
We then partition $U$ according to the recursion tree: each point $p\in U$ follows the cuts from the root to a leaf of $T$, according to whether it lies inside or outside each ball, and is assigned to the piece corresponding to that leaf.
For each final part, we choose the center of its $\tau$-ball as its representative.
For each other nonempty piece $U'$, if $U'$ fits on one machine, we gather $U'$ onto a single machine and repeatedly choose an arbitrary remaining point $z\in U'$ and remove the points in $B(z,\tau)$ from $U'$ as a final part with representative $z$.
Otherwise, we recursively process $U'$ in the next phase.

After $O(1/\sigma)$ phases, report failure if there is a remaining piece.
Otherwise, the final parts form a partition of $X$, with each part having a representative.
Let $X'$ be the set of all such representatives, and let $f(p)$ be the representative of the part containing $p$. By construction, $\dist(p,f(p))\leq \tau$ for every $p\in X$.

\paragraph{Analysis.}
This algorithm is a simplified version of the algorithm in~\cite{oh_et_al:LIPIcs.ESA.2026.88}. Their analysis immediately implies that our construction uses $O(1/\sigma^2)$ rounds and $O(n)$ total space.
We analyze the error induced by the perturbation.
The distance between two points increases by at most $2\tau$ since $\dist(p,f(p))\leq \tau$ for every $p\in X$ by construction, so $\opt(X')\leq\opt(X)+2n\tau$. Since $\opt(X)\geq \Delta=n^2\tau$, we have $\opt(X')\leq (1+o(1))\cdot \opt(X)$.
 
We show that the spread of $X'$ is $O(n^8)$. For two distinct points of $X'$, consider the first cut separating them. 
There are two cases.
In the first case, the cut is $B(z,\tau)$.
One representative is its center $z$ and the other lies outside $B(z,\tau)$, so their distance is greater than $\tau$.
In the second case, the cut is $B(z,r')$, where $r'$ is chosen uniformly from $[r,2r]$ with $r>\tau$.
Conditioned on the preceding choices, such a cut separates two fixed input points with probability at most their distance divided by $\tau$. There are $O(n)$ such cuts, since each splits a nonempty piece into two nonempty pieces. A union bound over these cuts and all input pairs shows that, with probability at least $1-O(n^{-3})$, no pair at distance less than $\tau/n^6$ is separated by a random cut. Together with the sampling guarantee, distinct representatives are therefore at distance at least $\tau/n^6$ with high probability. Their maximum distance is at most $2\Delta$, so the spread of $X'$ is $O(n^8)$. Its doubling dimension is at most $2d$.
Once we can compute a $1+\eps$-approximate MST of $X'$, then it yields a $1+o(1)+\eps$-approximate MST of $X$.

\section{Proof of Theorem~\ref{thm:doubling-mst} assuming Lemma~\ref{lem:balanced-light-partition-algorithm-doubling}}
\label{apd:doubling-proof}
We first describe our algorithm.
Let $L=\lceil 20/\sigma\rceil$, and keep $\EMPH{b}=n^{\sigma/10}$ fixed throughout the algorithm.
Starting with $P=Q_0$, we repeatedly compute a $(b,O(1))$-randomized light partition.
At level $i$, we are given a set $Q_i\subseteq P$ of representative points.
Let $(\mathcal C_i, \{T_C\}_{C\in \mathcal C_i})$ be a $(b,O(1))$-randomized light partition of $Q_i$ obtained by using Lemma~\ref{lem:balanced-light-partition-algorithm-doubling}.
We choose an arbitrary representative point per cluster in $\mathcal C_i$, and let $Q_{i+1}$ be the set of chosen representative points, which will be passed to the next level.
Once $n^{\sigma/2}$ representative points remain, we can construct their MST on a single machine.
On every successful call, the number of points becomes $O(m/b)+1$; while $m>n^{\sigma/2}$, this is at most $m/n^{\sigma/20}$ for sufficiently large $n$.
Thus $L$ successful calls suffice.
The union of the trees constructed at all levels forms a spanning tree on $P$. We denote it by $T$.


We now show that $T$ is an $O(1)$-approximate MST in expectation.
Suppose that there are $\ell<L$ levels.
For each $0\le i\le \ell$, let $m_i=|Q_i|$.
For every $i<\ell$, the trees constructed within the clusters in $\mathcal C_i$ form a spanning forest on $Q_i$, and $Q_{i+1}$ contains exactly one representative from each tree.
Therefore, adding the spanning forest of $Q_{i+1}$ merges those trees in the forest of $Q_i$ without creating a cycle, and thus $T$ is a spanning tree of $P$.
Note that $|E(T)|=\sum^{\ell-1}_{i=0}(\sum_{C\in \mathcal C_i}(|C|-1))+(m_\ell -1)=\sum^{\ell-1}_{i=0}(m_i-m_{i+1})+(m_\ell -1)=n-1$.

Consider an MST $T^*$ of $P$ and an Euler tour of the multigraph obtained by duplicating all edges of $T^*$.
The length of the Euler tour is $2\cdot \omega(T^*)$.
For each $0\le i\le \ell$, let $(q_1,q_2,\ldots ,q_{m_i})$ be the points in $Q_i$ ordered by their first occurrence in the Euler tour.
Then, $\sum_{j=1}^{m_i}\dist(q_j,q_{j+1})\leq 2\cdot \omega(T^*)$ by the triangle inequality, where $q_{m_i+1}=q_1$.
This implies that $\opt(Q_i)\leq 2\cdot \omega(T^*)=2\opt(P)$.
At each executed level, the second property of a randomized light partition bounds the expected total weight of the cluster trees by $O(\opt(Q_i))$, conditional on the preceding iterations and using fresh random choices at that level.
Thus, summing over the $O(1/\sigma)$ possible levels and including the final MST, we have
\[
\begin{aligned}
    \mathbb E[\omega(T)]&=\mathbb E\Bigl[\sum_{i=0}^{\ell-1}\sum_{C\in \mathcal C_i}\omega(T_C)+\opt(Q_\ell)\Bigr] \leq O((1/\sigma)\cdot \opt(P))+2\opt(P)=O((1/\sigma)\cdot \opt(P)).
\end{aligned}
\]

Each iteration constructs a randomized light partition in $O(\ddim/\sigma^3)$ MPC rounds.
Moreover, we can select representatives locally and update the representative and edge records in $O(1/\sigma)$ MPC rounds.
The MST on $Q_\ell$ is computed locally.
Since there are $\ell=O(1/\sigma)$ iterations, the algorithm takes $O(\ddim/\sigma^4)$ MPC rounds in total.

By the Markov inequality, the probability that $\omega(T)>2c\cdot \opt(P)$ is at most $1/2$ for some constant $c$.
Thus, by applying Algorithm~\ref{alg:MST-with-BLP} independently to $O(\log n)$ copies of $P$ in parallel and returning the tree of minimum total weight, we can construct an $O(1)$-approximate MST with high probability.
By reusing space across iterations, the bounds in Lemma~\ref{lem:balanced-light-partition-algorithm-doubling} also give $n^{\sigma}$ local space per machine, and $\widetilde{O}(n^{1+\sigma}\cdot 2^{O(\ddim)})$ total space.

\section{Proof of Lemma~\ref{lem:partitioned-tree-sum-local-rad}}
\label{apd:h-sum}

\euclideanLocalRad*
\begin{proof}
    It suffices to show that the right-hand side of Equation~(\ref{equ:euclidean-low-cost}) is $O(\opt(P))$.
    The second term of the right-hand side of the equation is $O(\opt(P))$ since $F$ is contained in the MST of $P$.
    Hence, we focus on proving $\sum_{C\in\textsc{cc}(F)}\min_{v\in C}h_v=O(\opt(P))$.
    
    If $F$ has a single component, then $C=P$, and we have $\min_{v\in P}h_v\leq \frac{\varepsilon\gamma}{50}\cdot \max_{x,y\in P}\dist(x,y) =O(\opt(P))$ by Lemma~\ref{lem:local-radius}.
    Otherwise, let $T$ be the MST of $P$ from Prim's algorithm.
    For a component $C$ of $F$, we can choose an edge $(u,v)\in E(T)$ with $u\in C$ and $v\in P\setminus C$.
    If $v\notin S_u$, $h_u\leq \varepsilon\hat r_u/100<\dist(u,v)/100$.
    Consider the call \textsc{TruncatedPrim}$(u,S_u,\varepsilon\hat r_u/2,0)$.
    The tree visited by this call is contained in $F[C]$.
    There are three cases when the subroutine terminates: either the total weight of the edges and vertices visited by the call is at least $\varepsilon\hat r_u/2$, the next edge has length at least $\varepsilon\hat r_u/2$, or all vertices in $S_u$ are visited.
    In the first case, $\omega(F[C])\geq \varepsilon\hat r_{u}/2$, and thus $50\cdot h_{u}\leq \varepsilon\hat r_{u}/2\leq \omega(F[C])$ by Lemma~\ref{lem:local-radius}.
    In the second case, $50\cdot h_{u}\leq \varepsilon\hat r_{u}/2\leq \dist(u,v)$.
    In the third case, $\dist(u,v)>\varepsilon\hat r_u$ since $B(u,\varepsilon\hat r_u)\cap P\subseteq S_u$.
    Thus, in all four cases, $50\cdot h_{u}\leq \omega(F[C])+\dist(u,v)$.
    We then have
    \[
    \begin{aligned}
        \sum_{C\in \textsc{cc}(F)}\min_{v\in C}h_v\leq \frac{1}{50}\sum_{C\in \textsc{cc}(F)}\left(\omega(F[C])+\dist(u_C,v_C)\right)\leq \frac{1}{50}(\omega(F)+2\omega(T))=O(\opt(P)),
    \end{aligned}
    \]
    where $(u_C,v_C)$ is the edge in $T$ chosen for $C$.
    The second inequality holds since each edge $(u,v)$ of $T$ is chosen in at most two components of $F$: one containing $u$ and one containing $v$.
\end{proof}

\end{document}